\documentclass[aps,prb,reprint,showpacs,superscriptaddress,longbibliography]{revtex4-2}

\usepackage[dvipdfmx]{graphicx}

\usepackage{floatrow}
\usepackage{here}
\usepackage{subcaption}

\DeclareCaptionLabelFormat{myformat}{#2}
\usepackage{amsmath,amssymb,amsfonts}
\usepackage{physics}

\usepackage[T1]{fontenc}
\usepackage{textcomp}
\usepackage{hyperref}
\hypersetup{colorlinks=true, linkcolor=blue, citecolor=blue, urlcolor=blue}
\DeclareUnicodeCharacter{2061}{}

\DeclareMathOperator{\Pf}{Pf}

\begin{document}

\title{Symplectic-Amoeba formulation of the non-Bloch band theory for one-dimensional two-band systems}

\author{Shin Kaneshiro}
\email{kaneshiro.shin.88a@st.kyoto-u.ac.jp}
\author{Robert Peters}
\affiliation{Department of Physics, Kyoto University, Kyoto 606-8502, Japan}
\date{\today}

\begin{abstract}
The non-Hermitian skin effect is a topological phenomenon, resulting in the condensation of bulk modes near the boundaries.
Due to the localization of bulk modes at the edges, boundary effects remain significant even in the thermodynamic limit.
This renders conventional Bloch band theory inapplicable and hinders the accurate computation of the spectrum.
The Amoeba formulation addresses this problem by determining the potential from which the spectrum can be derived using the generalized Szeg\"o's limit theorem,
reducing the problem to an optimization of the Ronkin function.
While this theory provides novel insights into non-Hermitian physics, challenges arise from the multiband nature and symmetry-protected degeneracies, even in one-dimensional cases.
In this work, we investigate one-dimensional two-band class AII$^\dagger$ systems, where Kramers pairs invalidate the conventional Amoeba formalism.
We find that these challenges can be overcome by optimizing the symmetry-decomposed Ronkin functions, which is achieved by extrapolating the total Ronkin function.
Finally, we propose a generalized Szeg\"o's limit theorem for class AII$^\dagger$ and numerically demonstrate that our approach correctly computes the potential and localization length.
\end{abstract}

\maketitle

\section{Introduction}
Non-Hermitian matrices are ubiquitous in a wide variety of physical systems, from classical wave dynamics to open quantum systems \cite{Ashida-Non-hermitianPhysics-2020e, Bergholtz-ExceptionalTopologySystems-2021u, Ding-Non-HermitianTopologyGeometries-2022h}.
These matrices naturally describe non-conservative dynamics induced by dissipation and energy exchange in real systems.
The presence of non-Hermiticity profoundly affects system properties, often leading to singular spectral structures with unique eigenstate distributions.
A particularly striking consequence of non-Hermiticity is the Non-Hermitian Skin Effect (NHSE) 
\cite{Yao-EdgeStatesSystems-2018f, Kunst-BiorthogonalBulk-BoundarySystems-2018e,
Yokomizo-Non-BlochBandSystems-2019i, Yang-Non-HermitianBulk-boundaryTheory-2019u, Kawabata-SymmetryAndPhysics-2019h, Lee-AnatomyOfSystems-2019s, Kunst-Non-HermitianSystemsPerspective-2019i, Song-Non-HermitianTopologicalSpace-2019r,
Zhang-CorrespondenceBetweenSystems-2020p, Okuma-TopologicalOriginEffects-2020t, Li-CriticalNon-HermitianEffect-2020k,  Kawabata-Higher-orderNon-HermitianEffect-2020k, Kawabata-Higher-orderNon-HermitianEffect-2020k, Longhi-Non-Bloch-BandCollapseTunneling-2020e, Yi-Non-HermitianSkinEffect-2020e,
Yokomizo-ScalingRuleEffect-2021u, Okugawa-Non-HermitianBandSymmetry-2021p, Lu-MagneticSuppressionEffects-2021y,
Wu-ConnectionsBetweenSystems-2022u, Liang-DynamicSignaturesAtoms-2022w, Song-Non-BlochPTSurprise-2022k, Longhi-Non-HermitianSkinSelf-accelerati-2022s, Gu-TransientNon-HermitianEffect-2022a,
Kawabata-EntanglementPhaseEffect-2023q, Li-EnhancementOfFields-2023n, Kawabata-EntanglementPhaseEffect-2023q,
Zhang-EdgeTheoryDimensions-2024x, Zhang-HybridSkin-topologicaNumbers-2024r}.
This phenomenon, characterized by a point-gap spectral structure, arises from the interplay between non-Hermiticity and topology, which does not have a counterpart in Hermitian systems.

The NHSE fundamentally modifies the system's response to boundary conditions, resulting in exponentially localized bulk modes and a breakdown of conventional Bloch band theory.
As a result, the eigenvalues and eigenstates under open boundary conditions (OBC) significantly deviate from those under periodic boundary conditions (PBC).

Complex wavenumbers with non-zero imaginary parts have been introduced to overcome this limitation, leading to a modified Bloch band theory.
These generalizations are known as the non-Bloch band theory, which has established the concept of the generalized Brillouin zone
\cite{Yao-EdgeStatesSystems-2018f, Yokomizo-Non-BlochBandSystems-2019i}.
This non-Bloch band theory has achieved remarkable success in one-dimensional systems, including symmetry-protected cases \cite{Kawabata-Non-BlochBandClass-2020k}, providing a comprehensive framework for understanding non-Hermitian one-dimensional systems
\cite{Longhi-Non-Bloch-BandCollapseTunneling-2020e,
Hu-KnotsAndBands-2021v, Yokomizo-Non-BlochBandSystems-2021g, Li-Non-BlochQuenchDynamics-2021e, Guo-Non-HermitianBulk-boundaryZone-2021c, Xue-SimpleFormulasTheory-2021y,
Li-TopologicalEnergyBands-2022x, Li-ExactFormulasSystems-2022p,
Hu-GreensFunctionsSystems-2023u, Hu-Non-BlochBandSystems-2023g, Liu-ModifiedGeneralizedDisorder-2023n, Hu-Non-BlochBandSystems-2023g, Tai-ZoologyOfTopology-2023l,
Matsushima-Non-BlochBandSystems-2024u, Verma-Non-BlochBandPhases-2024q, Wang-Non-BlochSelf-energyFermions-2024f, Roy-TopologicalCharacterizatioTheory-2024i, Yang-EntangelmentEntropyZone-2024p, Fu-BraidingTopologyBands-2024m, Yang-EntangelmentEntropyZone-2024g, Wang-GeneralTheorySystems-2024k, Hu-GeometricOriginBreaking-2024t, Matsushima-Non-BlochBandSystems-2024u, Verma-Non-BlochBandPhases-2024q,
Li-Phase-spaceGeneralizedSystems-2025e}.

In contrast, attempts to extend the non-Bloch band theory to higher dimensions have proved challenging 
\cite{Yao-Non-hermitianChernBands-2018x, Liu-Second-OrderTopologicalSystems-2019c, Yokomizo-Non-BlochBandsSystems-2023i, Jiang-DimensionalTransmutationNon-Hermiticity-2023f, Xu-Two-dimensionalAsymptoticConjecture-2023r, Zhang-AlgebraicNon-HermitianDimensions-2024f}.
While boundaries in one-dimensional systems are inherently simple, those in two or higher dimensions exhibit far more complex geometries.
In this context, the recently proposed Amoeba theory has significantly advanced the field
\cite{Ayan-ATropicalGeometricApproach-2023a,
Wang-AmoebaFormulationDimensions-2024q, Wang-ConstraintsOfSymmetry-2024w, Xiong-GraphMorphologyBands-2023o, Xiong-Non-HermitianSkinClassification-2024f, Hu-TopologicalOriginSpectra-2025i}.
The Amoeba formulation defines the "electrical" potential of the spectrum, making it possible to access the density of states (DOS) under OBC  for single-band class A systems in arbitrary dimensions.
In particular, the generalized Szeg\"o's limit theorem reduces the potential computation to an optimization of the Ronkin function, which is obtained by the PBC spectrum.
Furthermore, as the Ronkin function takes the inverse localization length as an argument, the solution to this optimization problem returns the correct localization length.

While the Amoeba formulation provides a powerful framework for developing the non-Bloch band theory of one-band systems,
generalizing it to multiband systems remains challenging.
In particular, degeneracies arising from the transpose-type time-reversal symmetry (TRS$^\dagger$) impose constraints on the Ronkin function \cite{Wang-ConstraintsOfSymmetry-2024w}, which obstruct both the optimization problem and the application of the generalized Szeg\"o's limit theorem. 
These constraints are especially significant in symplectic classes, where TRS$^\dagger$ leads to Kramer's degeneracies, further complicating the Amoeba formulation.
Consequently, even for one-dimensional systems, the Amoeba formulation fails to apply to symplectic classes.

Here, we partially overcome this limitation by generalizing the Amoeba formulation to the one-dimensional two-band class AII$^\dagger$.
Separating the Ronkin function into its symmetry-decomposed parts, which are related by the TRS$^\dagger$ operation,
we show that we can optimize these symmetry-decomposed Ronkin functions by extrapolating the total Ronkin function.
This extrapolation is easily calculated by the Legendre transformation of the total Ronkin function and leads to a generalized Szeg\"o's limit theorem for class AII$^\dagger$ in one dimension.

The rest of this paper is organized as follows:
In Sec. \ref{sec: OBC band for class A}, we introduce our notation and briefly review the non-Bloch band theory and the Amoeba formulation in one-dimensional class A systems.
In Sec. \ref{sec: GSLT for AIId}, we compare two fundamentally different approaches to decompose the Ronkin function, band-resolved and symmetry-decomposed Ronkin function, and demonstrate why only the latter preserves essential mathematical properties.
We then propose the generalized Szeg\"o's limit theorem for one-dimensional two-band class AII$^\dagger$ systems using the symmetry-based decomposition. 
In Sec. \ref{sec: Numerical verification}, we numerically verify our methods and theorems. 
Sec.~\ref{sec: Conclusions} summarizes and concludes this paper.
Furthermore, in Appendix~\ref{appendix: band-resolved Ronkin}, we provide a mathematical analysis demonstrating why the band-resolved Ronkin functions fail to inherit the mathematical properties from the total Ronkin function.
Appendix~\ref{appendix: GBZ from potent} derives the GBZ condition for class AII$^\dagger$ systems by analyzing the OBC spectral potential.

\section{OBC band structure for class A systems}
\label{sec: OBC band for class A}
This section introduces the notation and briefly reviews the non-Bloch band theory
\cite{Yao-EdgeStatesSystems-2018f, Yokomizo-Non-BlochBandSystems-2019i, Yang-Non-HermitianBulk-boundaryTheory-2019u}
and the Amoeba formulation
\cite{Wang-AmoebaFormulationDimensions-2024q}
in class A systems.

\subsection{Notations}
We consider a one-dimensional lattice of length $N$ with $M$ states at each site. For simplicity, we set the lattice constant to unity.
The OBC Hamiltonian $\hat{\mathcal{H}}$ is generally expressed as
\begin{align}
    \hat{\mathcal{H}} = \sum_{x,y=1}^N \sum_{a,b=1}^M \hat c_{x,a}^\dagger H^{(N)}_{xa,yb} \hat c_{y,b}
    \label{eq: OBC Hamiltonian}
\end{align}
where $\hat c_{x,a}, (\hat c_{x,a}^\dagger)$ annihilates (creates) a particle at site $x$ in state $a$.
Due to the periodicity of the lattice, the hopping amplitudes $H_{xa, yb}$ depend only on the relative distance between different sites and form a Block Toeplitz matrix $H^{(N)}$: $H^{(N)}_{xa,yb} = H_{x-y, ab}$.
The Fourier transformation of these matrix elements yields the Bloch Hamiltonian,
\begin{align}
    h_{\text{B},ab}(k) = \sum_{n=-p}^p H_{n, ab} e^{ikn},
    \label{eq: Bloch Hamiltonian}
\end{align}
where the parameter $p$ is the largest hopping distance.

In the absence of the NHSE, the spectrum of $H^{(N)}$ (OBC spectrum) is correctly approximated by the spectrum of $h_\mathrm{B}(k)$ (PBC spectrum) except for $\order{1}$ edge modes,
since the eigenstates are delocalized and approximated with plane waves with wavenumber $k$.
On the other hand, if the NHSE occurs, a large number $(\order{N})$ OBC eigenstates become exponentially localized, and complex-valued wavenumbers emerge.
To take into account such complex-valued wavenumbers, the non-Bloch Hamiltonian is defined through an analytic continuation of the Bloch Hamiltonian to $h_{\text{B}}(k-i\mu)$, where $\mu$ is the inverse localization length.
In the following discussion, we denote the non-Bloch Hamiltonian as 
\begin{align}
    h_\text{nB}(\beta) = h_\text{B} (- i \log \beta),
    \label{eq: non-Bloch Hamiltonian}
\end{align}
where $\beta=e^{\mu + ik}$.
In this $\beta$ representation, the non-Bloch Hamiltonian becomes a matrix-valued Laurent polynomial.

\subsection{Non-Bloch band theory}
The non-Bloch band theory determines the OBC band structure through a bivariate polynomial called the characteristic polynomial:
\begin{align}
    \mathrm{ChP}(E, \beta) = \det[E-h_{\text{nB}} (\beta)].
\end{align}
This polynomial establishes the relation between complex-valued wavenumbers $\beta$ and energies $E$ and generalizes the Bloch band theory.

The characteristic polynomial can be factorized in two complementary ways:
\begin{align}
    \mathrm{ChP}(E, \beta) &= C_E \beta^{-Mp} \prod_{j=1}^{2Mp} \qty(\beta - \beta_j(E)) \nonumber \\
    &= \prod_{\sigma = 1}^M \qty(E-E_\sigma(\beta))
    \label{eq: factorization of ch}
\end{align}
where $C_E$ is a $\beta$-independent constant.
We assume that the $\beta$-roots, $\beta_j$, are ordered by their moduli:
$\abs{\beta_i(E)} \le \abs{\beta_j(E)}$ for $i \le j$.
Then, the $E$-roots, $E_\sigma(\beta)$, provide the energy band branches (EBB)  \cite{Fu-AnatomyOfSystems-2023p}.
For a model described by a rank-$M$ matrix $h(\beta)$, there are $M$ EBBs.

There are several $\beta$ values associated with a given $E$ through the characteristic polynomial.
It has been found that for OBC, only a few $\beta_j(E)$ contribute to the OBC wave function.
The GBZ is defined by collecting these $\beta$ values.
In class A systems, the GBZ for band $\sigma$ is characterized as:
\begin{align}
    \text{GBZ}_\sigma = \{
        \beta \in \mathbb{C}
    ;
        \abs{\beta_{Mp} \circ E_\sigma(\beta)} 
        = \abs{\beta_{Mp+1} \circ E_\sigma(\beta)}
    \}
    \label{eq: GBZ criterion for class A},
\end{align}
where $\circ$ denotes a function composition \cite{Yao-EdgeStatesSystems-2018f, Yokomizo-Non-BlochBandSystems-2019i, Yang-Non-HermitianBulk-boundaryTheory-2019u}.
The spectrum of band $\sigma$ is then obtained by mapping the $\mathrm{GBZ}_\sigma$ through $E_\sigma$.

While the non-Bloch band formulation is applicable to multiband systems, generalizations to higher-dimensional systems are non-trivial since conditions analogous to Eq.~(\ref{eq: GBZ criterion for class A}) have not yet been established in higher-dimensional systems.

\subsection{Amoeba formulation}
The Amoeba formulation determines the OBC band structure through the OBC DOS and its potential.
The OBC DOS on the complex-energy plane is defined using the Dirac's delta function as
\begin{align}
    \rho(E) = \lim_{N \to \infty} \frac{1}{N} \Tr \delta \qty(E-H^{(N)}).
\end{align}
This DOS is intimately connected to the OBC spectral potential $\phi(E)$ which satisfies the Laplace equation
\begin{align}
    \rho(E) &= \frac{1}{2\pi} \Delta \phi(E),
    \label{eq: DOS and Potential}
\end{align}
where $\Delta = {\partial^2}/{\partial (\Re E)^2} + {\partial^2}/{\partial (\Im E)^2}$.
Since the complex-energy plane is two-dimensional, the potential can be rewritten as
\begin{align}
    \phi(E) = \lim_{N \to \infty} \frac{1}{N} \Tr \ln \abs{E-H^{(N)}}.
    \label{eq: potential 1}
\end{align}

While the direct calculation of this potential requires the determinant of a huge matrix,
it can be efficiently computed using the non-Bloch Hamiltonian under certain conditions.

In one-band class A systems, the spectral potential is given by Szeg\"o's limit theorem 
\cite{Szego-EinGrenzwertsatzFunktion-1915w},
\begin{align}
    \phi(E) = \oint_{\ln \abs{\beta} = 0} \frac{d \beta}{2\pi i \beta} \ln \abs{\mathrm{ChP}(\beta, E)},
    \label{eq: Szego limit theorem}
\end{align}
for topologically trivial $E$, detected by a zero-winding number $W(E)=0$, defined as
\begin{align}
    W(E) = \int_0^{2\pi} \frac{dk}{2\pi i} \partial_k \ln \det [E - h_{\mathrm B}(k)].
    \label{eq: Z topo. inv.}
\end{align}
This theorem states that the PBC spectral potential can be used for the OBC potential for trivial $E$.

Szeg\"o's limit theorem is invalid for topologically nontrivial $E$, though a generalization is proposed in \cite{Wang-AmoebaFormulationDimensions-2024q}.
According to this conjecture, the potential $\phi(E)$ in one-dimensional single-band systems $(M=1)$ is given by a test potential 
 $\Phi(E)$, which is determined through the following optimization problem:
\begin{align}
    \Phi(E) &= \min_\mu R_{E} (\mu),
    \label{eq: generalized Szego limit theorem class A}
\end{align}
where $R_E$ is the Ronkin function defined as
\begin{align}
     R_E (\mu) = \oint_{\ln\abs{\beta}=\mu} \frac{d \beta}{2\pi i \beta}
                 \ln \abs{ \mathrm{ChP}(E, \beta) }.
     \label{eq: Ronkin function}
\end{align}

The integration in Eq.~(\ref{eq: Ronkin function}) can be rewritten in terms of $\beta$ using Eq.~(\ref{eq: factorization of ch}):
\begin{align}
    R_E(\mu) &= \ln \abs{C_E} - p \oint_{\ln \abs{\beta}=\mu} \frac{d\beta}{2\pi i \beta} \ln \abs{\beta} \nonumber \\
    &+ \sum_{j=1}^{2p} \oint_{\ln \abs{\beta}=\mu} \frac{d\beta}{2\pi i \beta} \ln \abs{ \beta - \beta_j(E) }.
    \label{eq: Ronkin function in beta}
\end{align}
This representation indicates that the Ronkin function is determined by the pole and the roots of the characteristic equation enclosed by the circle $\abs{\beta} = e^\mu$ \cite{Xiong-GraphMorphologyBands-2023o}:
\begin{align}
    R_E(\mu) = \ln |C_E| - p\mu + \sum_{j=1}^{2p} \max(\mu_j, \mu)
    \label{eq: Ronkin function after integrate for A}
\end{align}
where $\mu_j=\ln|\beta_j(E)|$ and $\max(x, y)$ is the maximum of $x$ and $y$.
Thus, the Ronkin function is a convex and piece-wise linear function, and its derivative is quantized to integer values in one-dimensional systems.

Using this representation, we can expand the Ronkin function around its minimum as
\begin{align}
    R_E(\mu) = a + 
    \begin{cases}
        - (\mu - \mu_{p}) & \mu_{p-1} \le \mu \le \mu_{p} \\
        0 & \mu_{p\phantom{-1}} \le \mu \le \mu_{p+1} \\
        \mu - \mu_{p+1} & \mu_{p+1} \le \mu \le \mu_{p+2}
    \end{cases}
    ,
\end{align}
where $a$ is obtained by
\begin{align}
    a = \ln \abs{C_E} + \sum_{j=p+1}^{2p} \mu_j.
\end{align}

In the Amoeba formulation, the absence of an intermediate region where the Ronkin function is constant indicates that $E$ lies within the spectrum,
and the $\mu^*$, which minimizes the function, yields the corresponding eigenstate's inverse localization length.
This criterion, known as "central hole closing," is consistent with the GBZ condition in Eq.~(\ref{eq: GBZ criterion for class A}).
The Ronkin function can similarly be defined in higher-dimensional systems.
Hence, the Amoeba formulation provides a way to generalize non-Bloch band theory to higher-dimensional systems.

\section{Generalized Szeg\"o's limit theorem for two-band AII$^\dagger$ systems}
\label{sec: GSLT for AIId}
In this section, we focus on one-dimensional two-band class AII$^\dagger$ systems, which exhibit Kramers degeneracies protected by TRS$^\dagger$.
We show that these symmetry-protected degeneracies render the conventional Amoeba formalism inadequate.
We introduce a symmetry-decomposed Ronkin function that resolves the total spectral contribution into components corresponding to each Kramers partner distributed across the two bands.
We then demonstrate how symmetry-decomposed Ronkin functions can be optimized by extrapolating the total Ronkin function under symmetry constraints.
This approach leads to a modified version of the generalized Szeg\"o's limit theorem.

\subsection{Non-Bloch band theory for class-AII$^\dagger$ systems}
Systems in class AII$^\dagger$ are characterized by the following $\mathbb{Z}_2$ topological invariant \cite{Okuma-TopologicalOriginEffects-2020t}:
\begin{align}
    &(-1)^{\nu(E)} = \text{sgn} \left[
        \frac{\Pf[(E-h_{\text{B}}(\pi))T]}{\Pf[(E-h_{\text{B}}(0))T]}
        \right . \nonumber \\
        &\times \left .
         \exp \Bqty{
            -\frac{1}{2} \int_0^\pi dk \ \partial_k \ln \det[(E-h_{\text{B}}(k))T]
         }
        \right] .
        \label{eq: Z2 topo inv.}
\end{align}
The sign function, $\text{sgn}(x)$, takes the value $1$ for $x>0$ and $-1$ for $x<0$,
and $\Pf[A]$ corresponds to the Pfaffian of the skew-symmetric matrix $A$.
$T$ is a unitary operator fulfilling TRS$^\dagger$ symmetry defined as
\begin{align}
    T h_{\text{B}}^\top (k) T^{-1} = h_{\text{B}}(-k)\qc T T^* = -1.
    \label{eq: TRSd}
\end{align}
When expressed in terms of the non-Bloch Hamiltonian, this symmetry takes the form:
\begin{align}
    T h_{\text{nB}}^\top (\beta) T^{-1} = h_{\text{nB}}(\beta^{-1}).
\end{align}
Consequently, the characteristic polynomial remains invariant under the transformation $\beta \to \beta^{-1}$:
\begin{align}
    \mathrm{ChP}(\beta, E) = \mathrm{ChP}(\beta^{-1}, E).
\end{align}
This symmetry ensures that for each root $\beta(E)$, there exist its conjugate root $\beta^{-1}(E)$. Furthermore, the roots can be ordered as,
\begin{align}
    \abs{\beta_{2p}^{-1}} \le \dots \le \abs{\beta_{1}^{-1}} < 1 < \abs{\beta_{1}} \le \dots \le \abs{\beta_{2p}},
    \label{eq: roots ordering}
\end{align}
resulting in a modified GBZ condition \cite{Kawabata-Non-BlochBandClass-2020k}:
\begin{align}
    \text{GBZ}_\sigma = \{
        \beta \in \mathbb{C}
    ;
        \abs{\beta_{1} \circ E_\sigma(\beta)} = \abs{\beta_{2} \circ E_\sigma(\beta)}
    \}
    .
    \label{eq: GBZ criterion for class AIId}
\end{align}
These mathematical properties reflect the existence of Kramers pairs in the spectrum.

\subsection{Breakdown of the Amoeba formulation and band-resolved Ronkin function}

The symmetry properties of class-AII$^\dagger$ systems necessitate a modification of the conventional Amoeba formulation in Eq.~(\ref{eq: generalized Szego limit theorem class A}).

Since the Ronkin function becomes even \cite{Wang-ConstraintsOfSymmetry-2024w}, $\mu^*=0$ always minimizes the Ronkin function.
This seemingly implies that all bulk modes are delocalized, and the NHSE is absent in class-AII$^\dagger$ systems.
However, this conclusion contradicts both theoretical analyses and numerical results, which consistently demonstrate the existence of localized bulk modes and the NHSE in such systems \cite{Okuma-TopologicalOriginEffects-2020t, Kawabata-Non-BlochBandClass-2020k, Wang-ConstraintsOfSymmetry-2024w}.

For class AII$^\dagger$ systems, we can calculate the Ronkin function in Eq.~(\ref{eq: Ronkin function}) as
\begin{align}
    R_E(\mu) = \ln |C_E| - 2p\mu + \sum_{j=1}^{2p} [\max(-\mu_j, \mu) + \max(\mu_j, \mu)]
    \label{eq: Ronkin function after integrate for AII}
\end{align}
Note that all $\mu_j$ are positive as required by Eq.~(\ref{eq: roots ordering}).
By combining Eq.~(\ref{eq: Ronkin function after integrate for AII}) with the GBZ condition Eq.~(\ref{eq: GBZ criterion for class AIId}), we find that when $E$ lies within the spectrum, regions where the derivative of the Ronkin function equals $\pm 1$ disappear.
We refer to this criterion as "unit-slope hole closing."
This observation indicates that $\mu^*$ must be explicitly determined by these regions.

However, merely replacing $\mu^*$ in Eq.~(\ref{eq: generalized Szego limit theorem class A}) with the value determined by the jump of the derivative from $-2$ to $0$ and the jump from $0$ to $+2$, proves insufficient for accurately representing the true potential.
The source of this inaccuracy lies in the fact that the Ronkin function integrates contributions from both bands of the system:
\begin{align}
    R_E(\mu) = \sum_{\sigma = +, -} \oint_{\ln|\beta|=\mu} \frac{d \beta}{2\pi i \beta} \ln \abs{ E - E_\sigma(\beta) }.
\end{align}
Since this $\mu^*$ does not simultaneously minimize both bands, the obtained value for the potential differs from the true potential that would correctly include the contributions of both bands at their respective minima.

To address this issue, we can define band-resolved Ronkin functions for each band separately as
\begin{align}
    R^{[\sigma]}_{E} (\mu) = \oint_{\ln|\beta|=\mu} \frac{d \beta}{2\pi i \beta} \ln \abs{ E - E_\sigma(\beta) }.
    \label{eq: band-resolved Ronkin functions EBB}
\end{align}
We can expect that optimizing each function yields appropriate values for the potentials $\Phi_\sigma (E)$ and localization lengths $\mu_\sigma$ for each band.

This approach would seemingly resolve the optimization problems described above.
However, this straightforward definition of band-resolved Ronkin functions proves challenging.
The problem arises from the following critical issues:
These band-resolved Ronkin functions exhibit an explicit dependence on the branch cuts of the EBB.
Although Eq.~(\ref{eq: band-resolved Ronkin functions EBB}) can be used to calculate band-resolved Ronkin functions, it remains unclear which energy corresponds to each band-wise Ronkin function.
Simply using the eigenvalues of $h_\mathrm{nB}(\beta)$ results in a loss of essential mathematical properties. 
The resulting band-resolved Ronkin functions are neither convex nor piecewise linear, as demonstrated in Appendix~\ref{appendix: band-resolved Ronkin}.

These limitations necessitate an alternative approach that decomposes the Ronkin function according to its symmetry.
The following section presents a modified formulation that preserves the crucial mathematical properties while adequately accounting for the band structure.

\subsection{Constructing symmetry-decomposed Ronkin functions}

Szeg\"o's limit theorem, Eq.~(\ref{eq: Szego limit theorem}), holds when $\nu(E)$ is trivial. Therefore, we focus here on the case where $\nu(E)$ is topologically nontrivial.

First, we outline the mathematical properties that the symmetry-decomposed Ronkin function $R^{(\pm)}_E$ must have.
The symmetry-decomposed Ronkin functions should inherit the mathematical properties of the total Ronkin function $R_E$:
they must be convex, piece-wise linear, and have integer-quantized derivatives.
Furthermore, we require that $R^{(\pm)}_E$ exhibit central hole closing if $E$ is in the OBC spectrum, analogous to the class A case.
Considering Eq.~(\ref{eq: GBZ criterion for class AIId}), we assume that $R^{(+)}_E [R^{(-)}_E]$ remains constant in the domain $\mu_{1} \le \mu \le \mu_2$ $[-\mu_{2} \le \mu \le -\mu_{1}]$.

Next, we discuss how the TRS$^\dagger$ constrains the decomposed functions.
When a system satisfies TRS$^\dagger$, each state is doubly degenerate, forming Kramers pairs, and the states in each pair exhibit opposite localization lengths.
Consequently, we require the symmetry-decomposed Ronkin functions to have the same optimal value and to satisfy the following relation:
\begin{align}
R_E(\mu) = R_E^{(+)}(\mu) + R_E^{(-)}(\mu) \qc
R_E^{(+)}(\mu) = R_E^{(-)}(-\mu).
\label{eq: band-resolved Ronkin function and symmetry}
\end{align}

To construct such proper symmetry-decomposed Ronkin functions, we use Eq.~(\ref{eq: Ronkin function after integrate for AII}),
which shows that the total Ronkin function is determined by the $\beta$-roots and the poles enclosed by the circle $|\beta| = e^\mu$.
Therefore, we divide the roots and the order of the pole into two groups.
We assume that $R^{(+)}_E$ [$R^{(-)}_E$] contains the roots whose moduli are larger [smaller] than $1$.
From the above assumption, we can deduce that the $R^{(+)}_E [R^{(-)}_E]$ must remain constant in the domain $\mu_{1} \le \mu \le \mu_{2} [-\mu_2 \le \mu \le -\mu_1]$.
Thus, the order of poles must be divided into $1$ and $2p-1$.

Accordingly, the symmetry-decomposed Ronkin functions are given by,
\begin{align}
    R_E^{(+)}(\mu)
    &= \frac{1}{2} \ln \abs{C_E} - \frac{1}{2} \sum_{j=1}^{2p} \mu_j -\mu \nonumber \\
    &+ \sum_{j=1}^{2p} \max(\mu_j, \mu) \nonumber \\
    R_E^{(-)}(\mu) 
    &= \frac{1}{2} \ln \abs{C_E} + \frac{1}{2} \sum_{j=1}^{2p} \mu_j -(2p-1) \mu \nonumber \\
    &+ \sum_{j=1}^{2p} \max(-\mu_j, \mu)
    \label{eq: band-resolved Ronkin functions local}
\end{align}

Here, the second term provides a constant shift, ensuring that the symmetry-decomposed Ronkin functions satisfy Eq.~(\ref{eq: band-resolved Ronkin function and symmetry}).
The symmetry-decomposed Ronkin functions introduced here maintain the symmetry between bands while preserving the piecewise linearity and integer-valued derivatives of the Ronkin functions.
One can readily verify that the sum of these symmetry-decomposed Ronkin functions reproduces the total Ronkin function in Eq.~(\ref{eq: Ronkin function after integrate for AII}).

\subsection{Reconstruction of the total Ronkin function and generalized Szeg\"o's limit theorem}
We reconstruct the total Ronkin function using this decomposition and demonstrate that the extrapolation of the total Ronkin function optimizes the symmetry-decomposed Ronkin functions, leading to a generalized Szeg\"o's limit theorem for two-band class AII$^\dagger$ systems.

\begin{figure}[b]
    \centering
    \includegraphics[width=0.9\linewidth]{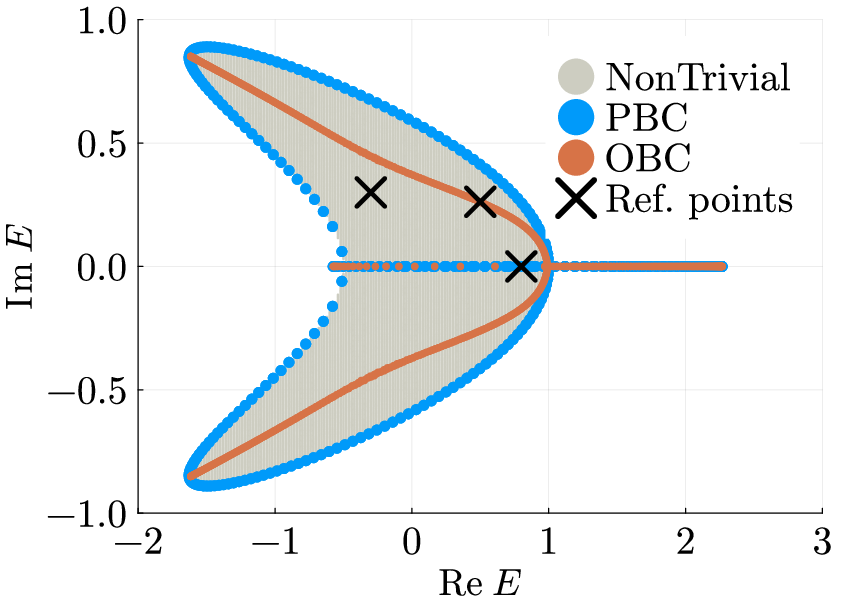}
    \caption{
        PBC and OBC ($N=400$) spectrum:
        The Hamiltonian is defined in Eq.~(\ref{eq:  Hamiltonian_benchmark}).
        Parameters are set to $t_1 = 0.3, t_2=0.8$, $g_{1,x} = 0.3, g_{1,z} = 0.5i$, $g_{2,x}=0.2$.
        The topologically nontrivial area is filled in gray.
        The reference points are $E=-0.3+0.3i, 0.5+0.265i, 0.8$.
    }
    \label{fig: benchmark_spectrum}
\end{figure}

\begin{figure*}[bt]
    \centering
    \begin{subfigure}{0.325\textwidth}
        \subcaption{}
        \vspace{-15pt}
        \includegraphics[width=0.85\linewidth]{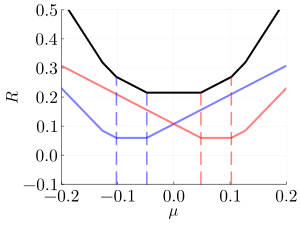}
    \end{subfigure}
    \hfill
    \begin{subfigure}{0.325\textwidth}
        \subcaption{}
        \vspace{-15pt}
        \includegraphics[width=.85\linewidth]{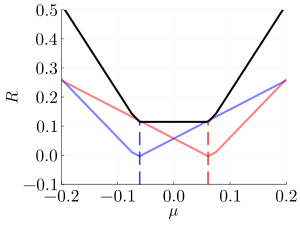}
    \end{subfigure}
    \hfill
    \begin{subfigure}{0.325\textwidth}
        \subcaption{}
        \vspace{-15pt}
        \includegraphics[width=0.85\linewidth]{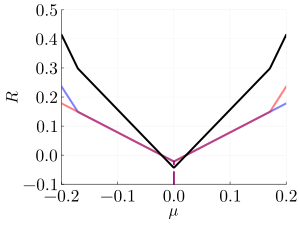}
    \end{subfigure}
    \caption{
        Total and symmetry-decomposed Ronkin functions evaluated at (a) $E=-0.3+0.3i$, (b) $E=0.5+0.265i$, and (c) $E=0.8$:
        The Hamiltonian and parameters are identical to those in Fig.~\ref{fig: benchmark_spectrum}.
        The black, red, and blue lines represent $R_E$, $R^{(+)}_E$ and $R^{(-)}_E$, respectively.
        The red and blue dashed lines indicate the central holes of $R_E^{(\pm)}$, which correspond to the unit-slope holes of $R_E$.
        The total Ronkin function is symmetric, and the symmetry-decomposed Ronkin functions $R^{(\pm)}$ satisfy the symmetry relation in Eq.~(\ref{eq: band-resolved Ronkin function and symmetry}).
        For points outside the OBC spectrum [(a)], both the central holes of $R_E^{(\pm)}$ and the unit-slope hole of $R_E$ open.
        For points inside the OBC spectrum [(b) and (c)], the unit-slope holes of $R_E$ vanish, accompanied by the closing of the central holes in the symmetry-decomposed Ronkin functions.
    }
    \label{fig: benchmark_partial_Ronkin_function}
\end{figure*}

The symmetry-decomposed Ronkin functions defined in Eq.~(\ref{eq: band-resolved Ronkin functions local}) can be expanded around their minima as
\begin{align}
    R^{(+)}_E(\mu) &= a + 
    \begin{cases}
        -(\mu - \mu_{1}) & -\infty \le \mu \le \mu_{1} \\
        0 & \phantom{-}\mu_{1} \le \mu \le \mu_{2} \\
        \mu - \mu_{2} & \phantom{-}\mu_{2} \le \mu \le \mu_3
    \end{cases} \\
    R^{(-)}_E(\mu) &= a + 
    \begin{cases}
        -(\mu + \mu_{2}) & -\mu_3 \le \mu \le -\mu_{2} \\
        0 & -\mu_{2} \le \mu \le -\mu_{1} \\
        \mu + \mu_{1} & -\mu_{1} \le \mu \le \infty
    \end{cases}
    .
\end{align} 
Here $a$ denotes the minimum of the symmetry-decomposed Ronkin functions,
\begin{align}
    a = \frac{1}{2} \ln \abs{C_E} - \frac{1}{2} \mu_1 + \frac{1}{2} \sum_{j=2}^{2p} \mu_j,
\end{align}
which corresponds to the potential of each band.

Because the total Ronkin function is symmetric, we focus on $\mu \ge 0$ and can easily reconstruct the total Ronkin function as
\begin{align}
    R_E(\mu) = 2a + 
    \begin{cases}
        2 \mu_1 & 0_{\phantom{1}} \le \mu \le \mu_1 \\
        \mu + \mu_1 & \mu_1 \le \mu \le \mu_2 \\
        2\mu - (\mu_2 - \mu_1) & \mu_2 \le \mu \le \mu_3
    \end{cases}.
\end{align}
We observe that the minimum of the Ronkin function is shifted by $2 \mu_1$ from the optimal value $2a$.
Therefore, we can obtain the true potential by correcting this error.
However, merely correcting this shift would result in the loss of hole-closing information, which plays a crucial role in the spectral detection for class A systems.
Thus, we focus on $\mu_2 - \mu_1$, which represents the length of the region with a slope of 1 and works as a unit-slope hole closing indicator.

With this knowledge, we can now extrapolate the total Ronkin function to obtain the correct potential.
By extrapolating the region with  a slope of $2$, we obtain the correct potential:
\begin{align}
   \Phi(E) = 2a = \tilde{R}_E^{(2)}(0) + (\mu_{2} - \mu_1),
   \label{eq: generalized Szego limit theorem class AIId extrapolate}
\end{align}
where $\tilde R^{(2)}_E(\mu) = 2\mu + 2a - (\mu_{2} - \mu_1)$.
Combining the GBZ condition Eq.~(\ref{eq: GBZ criterion for class AIId}), the extrapolated value yields the appropriate potential for the OBC spectrum, and the localization length is obtained by $\mu_1 (=\mu_2)$.

Notably, this extrapolation method works even when the region with a slope of $2$ vanishes. 
If $\mu_2=\mu_3$ holds, we can consider the tangent line with a slope of $2$ at $\mu=\mu_2$.
This condition motivates us to reformulate Eq.~(\ref{eq: generalized Szego limit theorem class AIId extrapolate}) in terms of the Legendre transformation.
 
We introduce the Legendre transformation of the Ronkin function, which takes the form
\begin{align}
    Q_E(m) = \min_{\mu} \qty[R_E(\mu) - m \mu].
\end{align}
In this framework, we can define the correct potential of the generalized Szeg\"o's limit theorem for two-band class AII$^\dagger$ systems as
\begin{align}
    \Phi(E) = Q_E(2)-(Q_E'(1+\epsilon)-Q_E'(1-\epsilon))
\end{align}
where $Q'_E$ denotes the derivative of $Q_E$, and $\epsilon$ represents an infinitesimal positive parameter
\footnote{
Since the Ronkin function is piecewise linear and has integer-quantized derivatives, the following equivalent expression exists:
$\Phi(E) = Q_E(2)-(Q_E'(2-\epsilon)-Q_E'(0+\epsilon))$.
}.

We can finally include the trivial case and summarize our conjecture to compute the OBC potential Eq.~(\ref{eq: potential 1}) as
\begin{align}
    \Phi(E) = 
    \begin{cases}
        Q_E(0) & \nu(E) = 0 \\
        Q_E(2) - (Q_E'(1+\epsilon) - Q_E'(1-\epsilon)) & \nu(E) = 1
    \end{cases}
    .
    \label{eq: generalized Szego limit theorem class AIId}
\end{align}
where $\nu(E)$ is the $\mathbb Z_2$ topological invariant defined in Eq.~(\ref{eq: Z2 topo inv.}).
The information about unit-slope hole closing in the total Ronkin function is inherited as the jump in the derivative $Q'_E$ at $m=1$.
Thus, the absence of such a jump serves as an indicator of the OBC spectrum.
When the unit-slope hole closes, the localization length $\mu^*(=\mu_1)$ is obtained by $Q_E^{-1}(1)$.
For more details, see Appendix~\ref{appendix: GBZ from potent}.

\section{Numerical verification}
\label{sec: Numerical verification}

Finally, in this section, we numerically demonstrate the correctness of our assumptions by calculating the potential and DOS using Eq.~(\ref{eq: generalized Szego limit theorem class AIId}).

We use the following Bloch Hamiltonian in class AII$^\dagger$,
\begin{align}
    h_{\mathrm B}(k) &= [2t_1 \cos k + 2t_2 \cos 2k] \sigma_0 \nonumber \\
    &- [2 \vb*g_1 \sin k + 2 \vb*g_2 \sin 2k] \cdot \vb*\sigma
    \label{eq: Hamiltonian_benchmark}
\end{align}
where $\sigma_0$ is the unit matrix, and $\sigma_{x,y,z}$ are the Pauli matrices.
The operator $T$ defined in Eq.~(\ref{eq: TRSd}) is given by $\sigma_y$.
The Hamiltonian can be diagonalized straightforwardly as,
\begin{align}
    E_{\pm}(k) = 2t_1 \cos k + 2t_2 \cos 2k
    \pm 2\norm{\vb*g_1 \sin k + \vb*g_2 \sin 2k}.
    \label{eq: EBB_benchmark}
\end{align}
We set the parameters as $t_1=0.3, t_2=0.8, \vb*g_1 = (0.3, 0.0, 0.5i), \vb*g_2=(0.2, 0.0, 0.0)$.

\begin{figure}
    \centering
    \includegraphics[width=\linewidth]{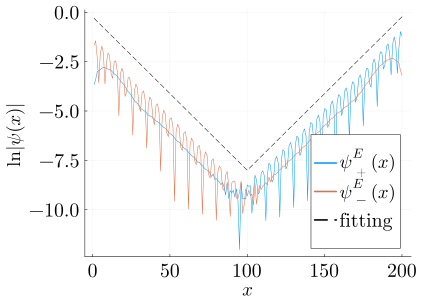}
    \caption{
    Fitting of wavefunctions using $\mu^* = Q_E^{-1}(1)$ for a system with $N=200$.
    The reference point is set to be $E=-0.26+0.43i$, which is inside the spectrum.
    The blue and orange lines represent the $+1$ and $-1$ sector of the Hamiltonian, respectively.
    Due to the symmetry, the wavefunction is exponentially localized at both boundaries, with a localization length of $1/\mu^*$.
    }
    \label{fig: fitting_wavefunction}
\end{figure}

\begin{figure}
    \centering
    \begin{subfigure}{\textwidth}
        \subcaption{}
        \vspace{-15pt}
        \includegraphics[width=0.85\linewidth]{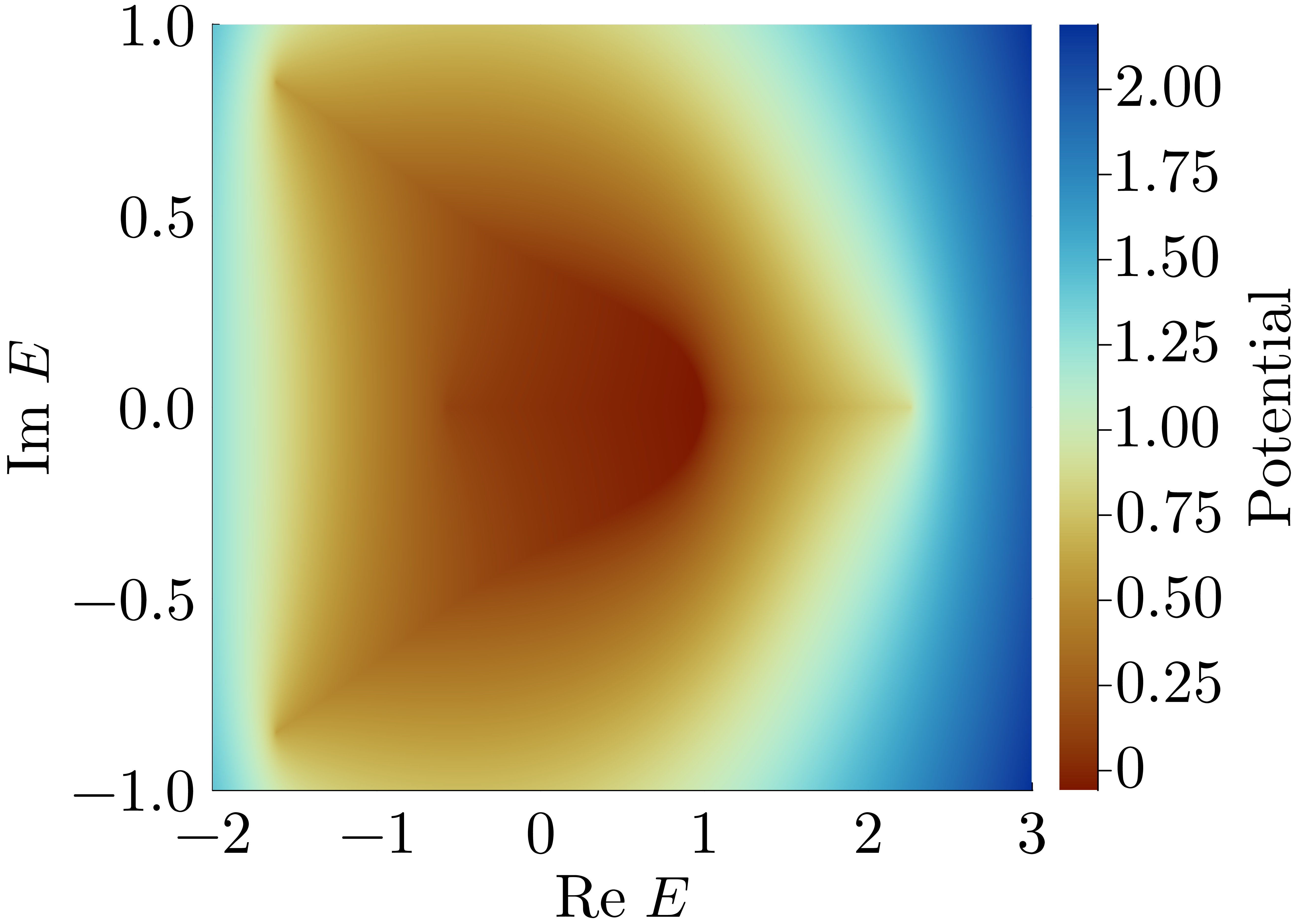}
    \end{subfigure}
    \\
    \begin{subfigure}{\textwidth}
        \subcaption{}
        \vspace{-15pt}
        \includegraphics[width=0.85\linewidth]{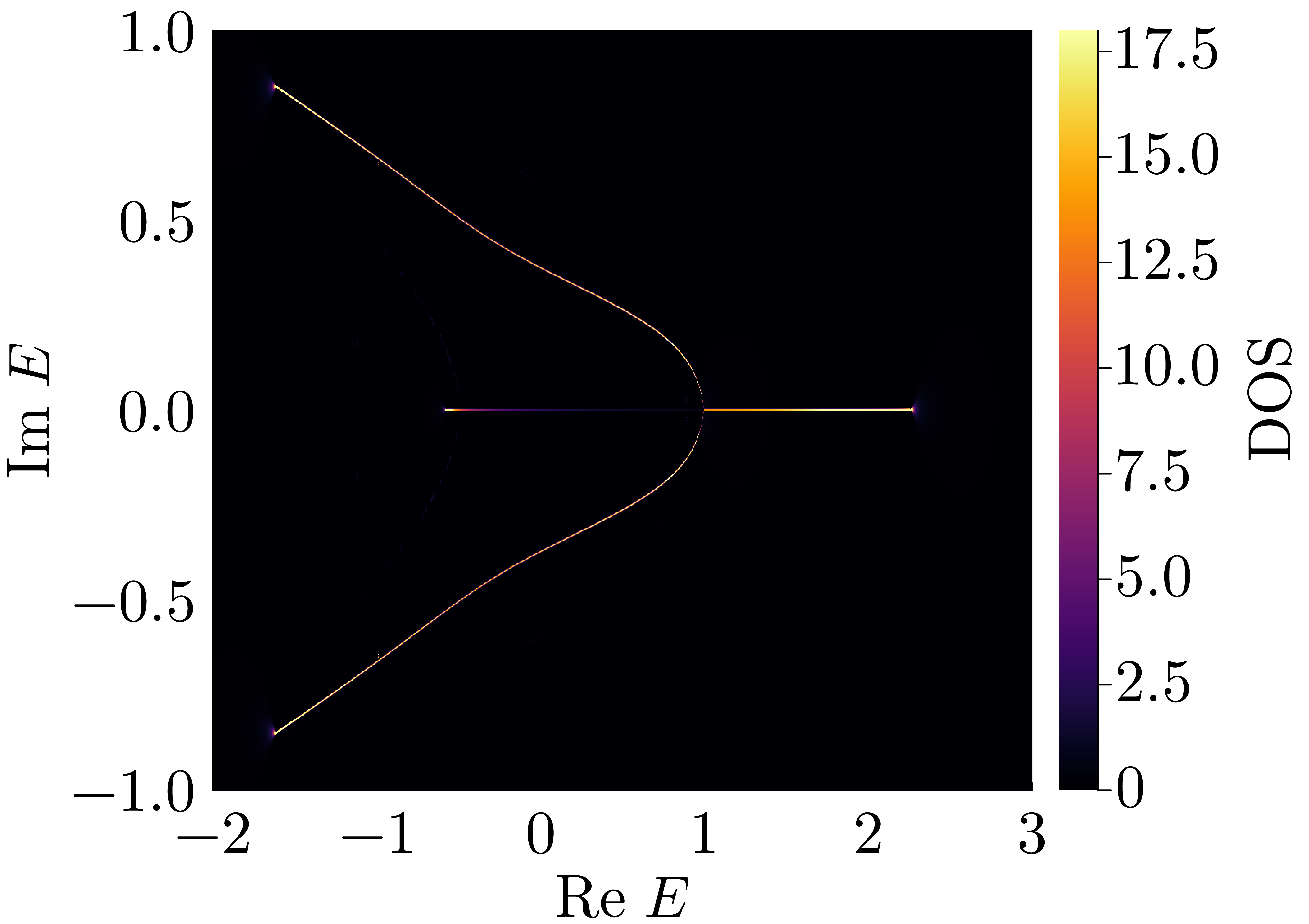}
    \end{subfigure}
    \caption{
        Spectral potential (a) and
        DOS given by this potential using Eq.~(\ref{eq: DOS and Potential}) in (b):
        The Hamiltonian and parameters are identical to those in Fig.~\ref{fig: benchmark_spectrum}.
        The direct diagonalization shows a $\psi$-shaped spectral distribution [Fig.~(\ref{fig: benchmark_spectrum})], which is successfully reproduced by our conjecture proposed in Eq.~(\ref{eq: generalized Szego limit theorem class AIId}).
    }
    \label{fig: Potential DOS benchmark}
\end{figure}

\begin{figure}
    \centering
    \includegraphics[width=0.9\linewidth]{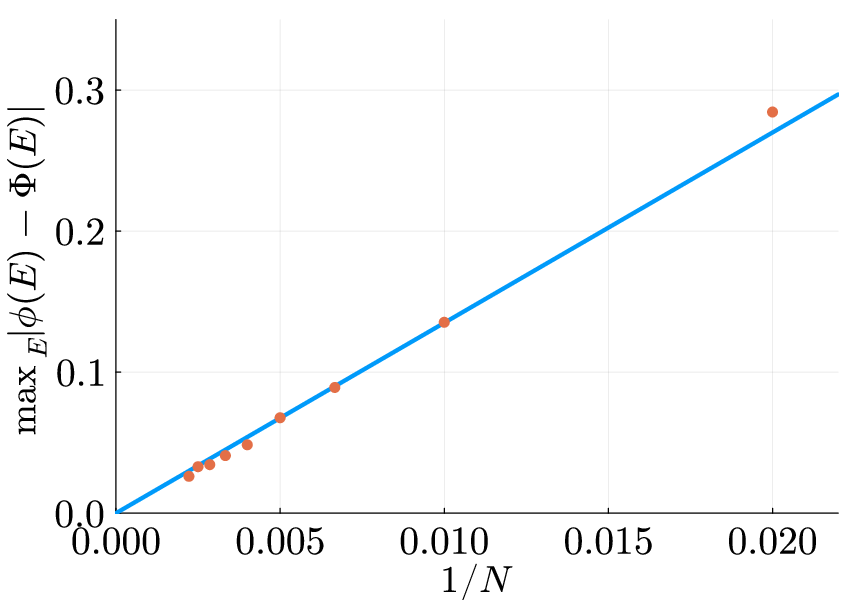}
    \caption{
        The maximum difference, $\max_E \abs{\phi(E)-\Phi(E)}$, between the OBC potential and the potential calculated by the Ronkin function for various system sizes $N=50, 100, \dots, 450$.
        The Hamiltonian and parameters are identical to those in Fig.~\ref{fig: benchmark_spectrum}.
        The difference converges to zero in the limit $N \to \infty$ with scaling $\order{N^{-1}}$.
    }
    \label{fig: Szego vs Ronkin class AIId}
\end{figure}

Figure~\ref{fig: benchmark_spectrum} shows the spectrum obtained by diagonalizing the finite-size Hamiltonian on a chain with 400 sites.
The blue and orange lines represent the spectra under PBC and OBC, respectively.
The complex plane contains a region with a nontrivial $\mathbb Z_2$ topological invariant, as defined in Eq.~(\ref{eq: Z2 topo inv.})  (shaded in gray).
The OBC spectrum coincides with the PBC spectrum along the real axis.
Both spectra match for $\Re E > 1$ but differ for $\Re E < 1$.
While the PBC spectrum is gapless, the OBC spectrum appears to exhibit a line gap at $E = 0.8$.
However, this gap arises due to finite-size effects and vanishes in the thermodynamic limit.

We show the total Ronkin functions and the symmetry-decomposed Ronkin functions in Fig.~\ref{fig: benchmark_partial_Ronkin_function}.
The reference points are set as $-0.3+0.3i$ [(a)], $E=0.5+0.265i$ [(b)], and $0.8$ [(c)].
In each figure, the black, red, and blue lines represent $R_E$, $R^{(+)}_E$, and $R^{(-)}_E$, respectively.
Due to TRS$^\dagger$ constraints, the total Ronkin function is symmetric around $\mu=0$.
The symmetry-decomposed Ronkin functions $R^{(\pm)}_E$ inherit the mathematical properties of the total function,
i.e., they are convex, piece-wise linear, and have integer-quantized derivatives.
Furthermore, they satisfy the symmetry relation in Eq.~(\ref{eq: band-resolved Ronkin function and symmetry}).
The red and blue dashed lines indicate the unit-slope holes of the total Ronkin functions inherited by the central holes of the symmetry-decomposed Ronkin functions.
For points outside the OBC spectrum [Fig.~\ref{fig: benchmark_partial_Ronkin_function}(a)], the central holes in the symmetry-decomposed Ronkin functions persist, which results in the opening of the unit-slope holes in the total Ronkin function.
For points inside the OBC spectrum [Fig.~\ref{fig: benchmark_partial_Ronkin_function}(b), (c)], the unit-slope holes in the total Ronkin function vanish,
which corresponds to the closing of the central holes in the symmetry-decomposed Ronkin functions.
At these points, the red and blue dashed lines mark $\pm\mu_1$, where $\mu_1=\mu^*$ denotes the inverse localization length of corresponding eigenstates.
These results are consistent with the GBZ condition in Eq.~(\ref{eq: GBZ criterion for class AIId}).

Figure~\ref{fig: fitting_wavefunction} compares an eigenstate of the finite-size system ($N=200$) with an exponentially localized wavefunction, using $\mu^* = Q_E^{-1}(1)$.
We set the reference point as $E=-0.26+0.43i$, which is located within the spectrum.
The blue [orange] line represents the $+1 [-1]$ sector of the Hamiltonian.
Due to the symmetry, the wavefunction is localized at both boundaries and exhibits linear behavior on a logarithmic scale, indicating exponential localization of the eigenstates with a localization length of $1/\mu^*$.

Figure~\ref{fig: Potential DOS benchmark} shows the potential and the DOS using Eq.~(\ref{eq: generalized Szego limit theorem class AIId}).
Our modified Amoeba formulation works in topologically nontrivial domains and successfully predicts the $\psi$-shaped spectral distribution,
consistent with numerical diagonalization results [Fig.~\ref{fig: benchmark_spectrum}].

Finally, to examine the generalized Szeg\"o's limit theorem, we perform finite-size analysis and calculate the maximum difference between the exact potential $\phi(E)$ (calculated by Eq.~(\ref{eq: potential 1})) and our proposed potential $\Phi(E)$ using Eq.~(\ref{eq: generalized Szego limit theorem class AIId extrapolate}), shown in Fig.~\ref{fig: Szego vs Ronkin class AIId}.
The difference converges to zero in the thermodynamic limit ($N \to \infty$) with scaling $\order{N^{-1}}$, which shows the validity of our conjecture.

\section{Conclusions}
\label{sec: Conclusions}
The Amoeba formulation has opened a new path for studying the GBZ in higher dimensions.
However, the multiband nature and symmetry-protected degeneracies invalidate the conventional Amoeba formulation since the optimization problem becomes invalid.

This paper partially solves this problem, focusing on one-dimensional two-band class AII$^\dagger$ systems.
Due to symmetry constraints based on TRS$^\dagger$, we can decompose the Ronkin functions into symmetry-decomposed functions and reconstruct the generalized Szeg\"o's limit theorem,
which can be calculated using the Legendre transformation.
We finally numerically demonstrate the correctness of our conjectures by calculating the spectrum of a two-band class AII$^\dagger$ system and showing the validity of the modified Szeg\"o's limit by comparing with exact diagonalization.

We note that a generalization of this extrapolation method to higher-dimensional cases remains challenging.
In one-dimensional systems, both the total Ronkin function and symmetry-decomposed Ronkin functions are piecewise linear, which helps recover the correct potential. 
However, the Ronkin function is no longer linear in higher-dimensional systems, making the extrapolation problematic.

\section*{Acknowledgments}
S.K. deeply appreciates the fruitful discussions with D. Nakamura, K. Shimomura, and K. Shinada.
This work is supported by JSPS, KAKENHI Grants No. JP24KJ1353 (S.K.), and No. JP23K03300 (R.P.).

\appendix
\section*{appendix}
\renewcommand{\thesubsection}{\Alph{subsection}}

\subsection{Band-resolved Ronkin functions}
\label{appendix: band-resolved Ronkin}

\begin{figure*}
    \centering
    \begin{subfigure}{0.245\textwidth}
        \subcaption{}
        \includegraphics[width=\linewidth]{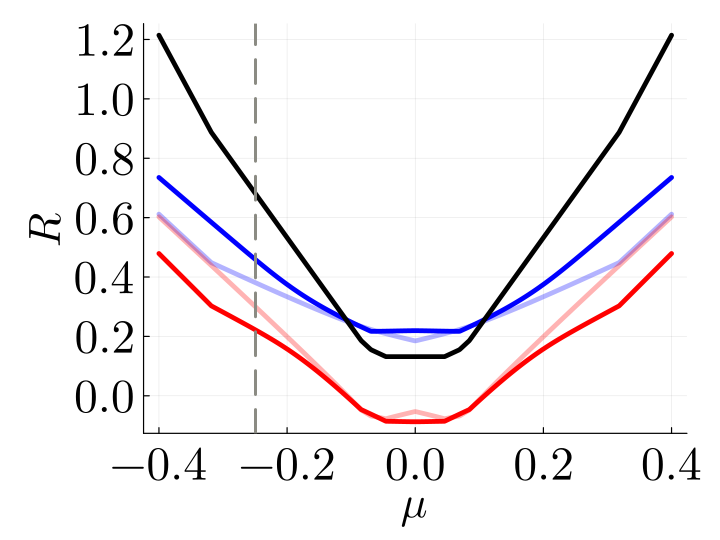}
    \end{subfigure}
    \hfill
    \begin{subfigure}{0.245\textwidth}
        \subcaption{}
        \includegraphics[width=\linewidth]{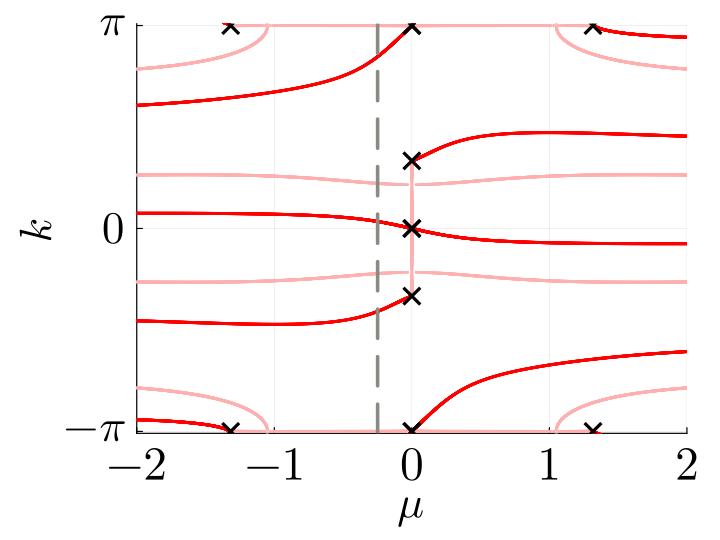}
    \end{subfigure}
    \hfill
    \begin{subfigure}{0.245\textwidth}
        \subcaption{}
        \includegraphics[width=\linewidth]{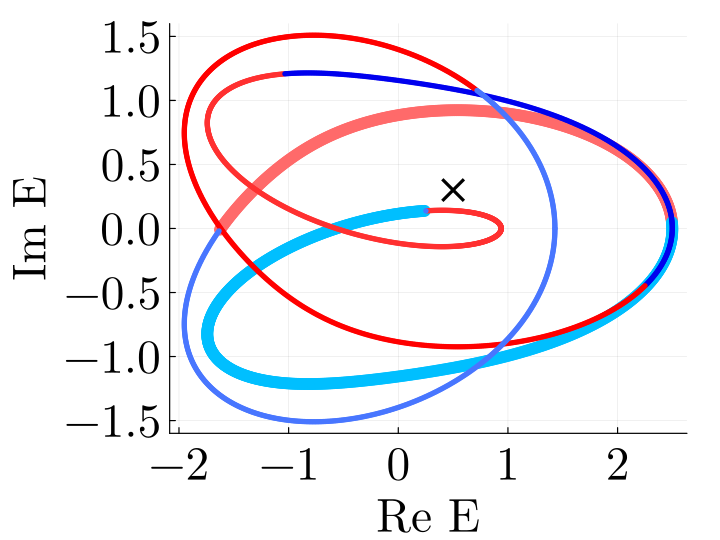}
    \end{subfigure}
    \hfill
    \begin{subfigure}{0.245\textwidth}
        \subcaption{}
        \includegraphics[width=\linewidth]{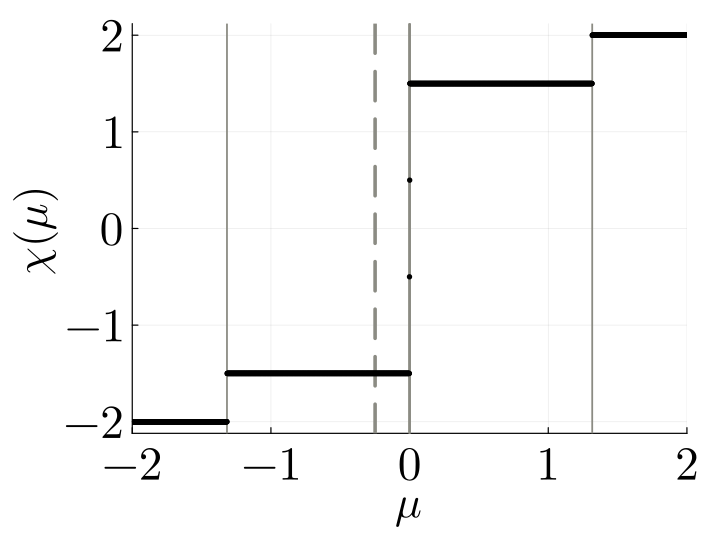}
    \end{subfigure}
    \caption{
    Breakdown of the mathematical properties of band-resolved Ronkin functions caused by branch-cut choice and eigenvalue braiding.
    (a) Band-resolved Ronkin functions $R_E^{[+]}$ (red) and $R_E^{[-]}$ (blue) at $E=0.5+0.3i$.
    Branch cuts of the square root extend from the origin at angles of $5\pi/7$ (red/blue) and $0$ (light red/light blue).
    Both branch cuts produce non-convex profiles with distinct minima, whereas their sum $R_E$ (black) is cut-independent and convex.
    The vertical dashed line marks $\mu=-0.25$ for reference across all panels.
    (b) The two branch cuts in the $(\mu, k)$-plane (red and light red), connecting the branch points (black $\times$).
    (c) Energy spectrum $E_\pm(\exp (\mu + ik))$ at $\mu=-0.25$ with branch cut $5\pi/7$.
    A single closed loop fractures into several open arcs (shown in different colors) due to the branch cuts.
    The black $\times$ marks the reference energy $E=0.5+0.3i$.
    (d) The band vorticity $\chi(\mu)$ as a function of $\mu$.
    Its value changes at the values of $\mu$ corresponding to the branch points shown in (b).
    For $\mu = -0.25$, $\chi(\mu)$ takes the half-integer value $-1.5$, which results in a single spectral loop as shown in (c).
    Consequently, no global branch choice can recover the closed spectral loops, resulting in a loss of convexity and integer-quantized slopes for the individual resolved functions.
    }
    \label{fig: EBB breakdown_of_convexity}
\end{figure*}

This appendix provides a mathematical analysis of why naive band-resolved Ronkin functions, defined in Eq.~(\ref{eq: band-resolved Ronkin functions EBB}), fundamentally fail to preserve the essential mathematical properties of convexity, piecewise linearity, and integer-quantized derivatives.
We demonstrate that the breakdown of these properties stems from the multi-valued nature of EBBs, which necessitates branch cuts that destroy the analytic structure underlying the Amoeba formulation.

To elucidate the fundamental difference between single-band and multiband systems, we first examine why the Ronkin function for single-band systems preserves the above properties.
Here, the EBB coincides with the single-band dispersion: $E(\beta) = h_\mathrm{nB}(\beta)$.
The derivative of the Ronkin function takes the form
\begin{align}
    \partial_\mu R_E(\mu) = \int_0^{2\pi} \frac{d k}{2\pi i} \partial_k \ln [E - E(e^{\mu + ik})],
\end{align}
which counts the winding of the curve consisting of $E(e^{\mu + ik})$ for $k \in [0, 2 \pi)$ around the reference energy $E$ \cite{Wang-AmoebaFormulationDimensions-2024q}.
Thus, the periodicity of the EBB directly ensures the piecewise linearity and integer-quantized derivatives of the Ronkin function.
Moreover, $E(\beta)$ is analytic on the whole complex plane except at the origin.
Consequently, the integrand $\ln |E-E(\beta)|$ is subharmonic, which ensures convexity of the Ronkin function \cite{Ronkin-Introduction}.
Thus, in the single-band case, both periodicity and analyticity are naturally satisfied, ensuring all desired mathematical properties.

In multi-band systems, the band-resolved Ronkin functions are constructed from the individual EBBs $E_\sigma(\beta)$ instead of $E(\beta)$.
Selecting a single branch from the eigenvalues of $h_\mathrm{nB}(\beta)$ requires the introduction of branch cuts, which fundamentally alter the analytic structure of the functions.
Branch cuts render the $E_\sigma(\beta)$ non-analytic along their loci;
As a result, the integrand $\ln |E-E_\sigma(\beta)|$ loses its subharmonic property in the vicinity of these cuts,
and the convexity of band-resolved Ronkin functions is no longer guaranteed.

Whether the derivative remains integer-quantized is dictated by the spectral braiding topology, which is encoded in the band vorticity
\cite{PhysRevLett.118.040401, PhysRevLett.120.146402, Li-TopologicalEnergyBands-2022x, PhysRevLett.126.086401}
\begin{align}
\chi(\mu) 
&= \frac{1}{2} \int_0^{2\pi} \frac{dk}{2\pi i}  \partial_k \ln \qty[E_+(e^{\mu+ik})-E_-(e^{\mu+ik})]^2 \nonumber\\
&= \frac{1}{2} \int_0^{2\pi} \frac{dk}{2\pi i}  \partial_k \ln \Delta(e^{\mu+ik})
\end{align}
where $\Delta = (\Tr h_\mathrm{nB})^2 - 4 \det h_\mathrm{nB}$ is the discriminant of the characteristic polynomial.
When $\chi(\mu)$ takes a half-integer value, the spectrum of $h_\mathrm{nB}$ forms a single closed curve, and the EBBs satisfy the band-swapping condition: $E_\pm(e^{\mu + i \cdot 0}) = E_\mp(e^{\mu + i \cdot 2\pi})$.
Consequently, no choice of branch cuts can simultaneously preserve both analyticity and $2\pi$-periodicity.
Under these conditions, some EBBs unavoidably trace open arcs rather than closed curves, destroying the integer-quantized derivatives of the band-resolved Ronkin function.

Therefore, while single-band Ronkin functions possess robust mathematical properties guaranteed by natural periodicity and analyticity, band-resolved functions in multi-band systems cannot offer the same guarantees due to the fundamental incompatibility between branch cut requirements and the preservation of these essential conditions.

Figure~\ref{fig: EBB breakdown_of_convexity} illustrates why the naive band-resolved Ronkin functions fail with numerical results.
The EBBs are defined in Eq.~(\ref{eq: EBB_benchmark}), and their parameters are set as in Fig.~\ref{fig: benchmark_spectrum}.
Figure~\ref{fig: EBB breakdown_of_convexity}(a) plots the total and the band-resolved Ronkin functions for $E=0.5+0.3i$.
The black, red and blue lines represent $R_E$, $R_E^{[+]}$ and $R_E^{[-]}$, respectively.
The EBBs include the square root of a complex number and, thus, depend on the branch cut.
We consider here two choices of branch cuts: a half-line from the origin to infinity at angles of $5\pi/7$ (red and blue) and 0 (light red and light blue).
The corresponding branch cuts on $(\mu, k)$-space are plotted in Fig.~\ref{fig: EBB breakdown_of_convexity}(b).
The partial functions $R_E^{[\pm]}$ have different minimal values, and their shapes strongly depend on the choice of branch cuts, although the total Ronkin function is cut-independent and convex.

The spectrum of $h_\mathrm{nB}$ with $\mu$ fixed at $-0.25$ forms a closed curve [Fig.~\ref{fig: EBB breakdown_of_convexity}(c)], resulting in the integer-quantized derivative in the total Ronkin function.
However, since the EBBs are non-analytic in the vicinity of the branch cuts, this non-analyticity causes the spectral curve to split into several open arcs.
Consequently, $R_E^{[\pm]}$ loses both its convexity and piecewise linearity, as demonstrated in Fig.~\ref{fig: EBB breakdown_of_convexity}(a).
As $\chi(\mu)$ takes half-integer values [Fig.~\ref{fig: EBB breakdown_of_convexity}(d)], it becomes impossible to partition the spectrum into closed curves regardless of the branch cut choice.
Consequently, the band-resolved Ronkin functions cannot recover their convexity and piecewise linearity.

These results demonstrate that any approach based on individual EBBs inevitably suffers from fundamental mathematical limitations.
This motivates the symmetry-based decomposition presented in the main text, which preserves the essential mathematical properties.

\subsection{Derivation of the GBZ criterion from spectral potential}
\label{appendix: GBZ from potent}
In this appendix, we derive the GBZ criterion for class AII$^\dagger$ systems given in Eq.(\ref{eq: GBZ criterion for class AIId}) by analyzing the OBC spectral potential calculated using Eq.(\ref{eq: generalized Szego limit theorem class AIId}).

The key insight is that the second term in Eq.~(\ref{eq: generalized Szego limit theorem class AIId}) quantifies the jump discontinuity in the derivative $Q_E'(m)$ at $m=1$.
As explained in the main text, this jump discontinuity directly encodes the information about unit-slope hole closing of the Ronkin function $R_E$.

To establish this connection quantitatively, we recall that the derivative of the Legendre transform returns the minimizing argument:
\begin{align}
Q_E'(m) = -\arg \min_\mu [R_E(\mu)-m\mu].
\end{align}
Using the explicit form of the Ronkin function for class AII$^\dagger$ systems from Eq.~(\ref{eq: Ronkin function after integrate for AII}), we can evaluate the left and right derivatives at $m=1$:
\begin{align}
Q_E'(1^-) &= \lim_{\epsilon \to 0^+} Q_E'(1-\epsilon) = -\mu_1, \\
Q_E'(1^+) &= \lim_{\epsilon \to 0^+} Q_E'(1+\epsilon) = -\mu_2,
\end{align}
where $\mu_1$ and $\mu_2$ are the boundary points of the unit-slope hole in $R_E(\mu)$.

The jump discontinuity $Q_E'(1^+) - Q_E'(1^-) = \mu_1 - \mu_2$ directly measures the width of the unit-slope hole in the Ronkin function. 
When this hole closes, the boundary points coincide ($\mu_1 = \mu_2$), eliminating the discontinuity.
Therefore, the closure condition $\mu_1 = \mu_2$ ensures continuity:
\begin{align}
Q_E'(1^-) = Q_E'(1^+) = -\mu_1 = -\mu_2,
\end{align}
which corresponds to the continuity of $Q_E'(m)$ at $m=1$.

This continuity condition is precisely the GBZ criterion $\mu_1 = \mu_2$ for class AII$^\dagger$ systems, thus establishing the equivalence between the spectral potential approach and the traditional GBZ formulation.

\bibliography{Symplectic_Amoeba}

\begin{thebibliography}{70}%
\makeatletter
\providecommand \@ifxundefined [1]{%
 \@ifx{#1\undefined}
}%
\providecommand \@ifnum [1]{%
 \ifnum #1\expandafter \@firstoftwo
 \else \expandafter \@secondoftwo
 \fi
}%
\providecommand \@ifx [1]{%
 \ifx #1\expandafter \@firstoftwo
 \else \expandafter \@secondoftwo
 \fi
}%
\providecommand \natexlab [1]{#1}%
\providecommand \enquote  [1]{``#1''}%
\providecommand \bibnamefont  [1]{#1}%
\providecommand \bibfnamefont [1]{#1}%
\providecommand \citenamefont [1]{#1}%
\providecommand \href@noop [0]{\@secondoftwo}%
\providecommand \href [0]{\begingroup \@sanitize@url \@href}%
\providecommand \@href[1]{\@@startlink{#1}\@@href}%
\providecommand \@@href[1]{\endgroup#1\@@endlink}%
\providecommand \@sanitize@url [0]{\catcode `\\12\catcode `\$12\catcode `\&12\catcode `\#12\catcode `\^12\catcode `\_12\catcode `\%12\relax}%
\providecommand \@@startlink[1]{}%
\providecommand \@@endlink[0]{}%
\providecommand \url  [0]{\begingroup\@sanitize@url \@url }%
\providecommand \@url [1]{\endgroup\@href {#1}{\urlprefix }}%
\providecommand \urlprefix  [0]{URL }%
\providecommand \Eprint [0]{\href }%
\providecommand \doibase [0]{https://doi.org/}%
\providecommand \selectlanguage [0]{\@gobble}%
\providecommand \bibinfo  [0]{\@secondoftwo}%
\providecommand \bibfield  [0]{\@secondoftwo}%
\providecommand \translation [1]{[#1]}%
\providecommand \BibitemOpen [0]{}%
\providecommand \bibitemStop [0]{}%
\providecommand \bibitemNoStop [0]{.\EOS\space}%
\providecommand \EOS [0]{\spacefactor3000\relax}%
\providecommand \BibitemShut  [1]{\csname bibitem#1\endcsname}%
\let\auto@bib@innerbib\@empty
\bibitem [{\citenamefont {Ashida}\ \emph {et~al.}(2020)\citenamefont {Ashida}, \citenamefont {Gong},\ and\ \citenamefont {Ueda}}]{Ashida-Non-hermitianPhysics-2020e}%
  \BibitemOpen
  \bibfield  {author} {\bibinfo {author} {\bibfnamefont {Y.}~\bibnamefont {Ashida}}, \bibinfo {author} {\bibfnamefont {Z.}~\bibnamefont {Gong}},\ and\ \bibinfo {author} {\bibfnamefont {M.}~\bibnamefont {Ueda}},\ }\bibfield  {title} {\bibinfo {title} {Non-hermitian physics},\ }\href {https://doi.org/10.1080/00018732.2021.1876991} {\bibfield  {journal} {\bibinfo  {journal} {Adv. Phys.}\ }\textbf {\bibinfo {volume} {69}},\ \bibinfo {pages} {249} (\bibinfo {year} {2020})}\BibitemShut {NoStop}%
\bibitem [{\citenamefont {Bergholtz}\ \emph {et~al.}(2021)\citenamefont {Bergholtz}, \citenamefont {Budich},\ and\ \citenamefont {Kunst}}]{Bergholtz-ExceptionalTopologySystems-2021u}%
  \BibitemOpen
  \bibfield  {author} {\bibinfo {author} {\bibfnamefont {E.~J.}\ \bibnamefont {Bergholtz}}, \bibinfo {author} {\bibfnamefont {J.~C.}\ \bibnamefont {Budich}},\ and\ \bibinfo {author} {\bibfnamefont {F.~K.}\ \bibnamefont {Kunst}},\ }\bibfield  {title} {\bibinfo {title} {Exceptional topology of non-hermitian systems},\ }\href {https://doi.org/10.1103/RevModPhys.93.015005} {\bibfield  {journal} {\bibinfo  {journal} {Rev. Mod. Phys.}\ }\textbf {\bibinfo {volume} {93}},\ \bibinfo {pages} {015005} (\bibinfo {year} {2021})}\BibitemShut {NoStop}%
\bibitem [{\citenamefont {Ding}\ \emph {et~al.}(2 12)\citenamefont {Ding}, \citenamefont {Fang},\ and\ \citenamefont {Ma}}]{Ding-Non-HermitianTopologyGeometries-2022h}%
  \BibitemOpen
  \bibfield  {author} {\bibinfo {author} {\bibfnamefont {K.}~\bibnamefont {Ding}}, \bibinfo {author} {\bibfnamefont {C.}~\bibnamefont {Fang}},\ and\ \bibinfo {author} {\bibfnamefont {G.}~\bibnamefont {Ma}},\ }\bibfield  {title} {\bibinfo {title} {Non-hermitian topology and exceptional-point geometries},\ }\href {https://doi.org/10.1038/s42254-022-00516-5} {\bibfield  {journal} {\bibinfo  {journal} {Nature Reviews Physics}\ }\textbf {\bibinfo {volume} {4}},\ \bibinfo {pages} {745} (\bibinfo {year} {2022-12})}\BibitemShut {NoStop}%
\bibitem [{\citenamefont {Yao}\ and\ \citenamefont {Wang}(2018)}]{Yao-EdgeStatesSystems-2018f}%
  \BibitemOpen
  \bibfield  {author} {\bibinfo {author} {\bibfnamefont {S.}~\bibnamefont {Yao}}\ and\ \bibinfo {author} {\bibfnamefont {Z.}~\bibnamefont {Wang}},\ }\bibfield  {title} {\bibinfo {title} {Edge states and topological invariants of non-hermitian systems},\ }\href {https://doi.org/10.1103/PhysRevLett.121.086803} {\bibfield  {journal} {\bibinfo  {journal} {Phys. Rev. Lett.}\ }\textbf {\bibinfo {volume} {121}},\ \bibinfo {pages} {086803} (\bibinfo {year} {2018})}\BibitemShut {NoStop}%
\bibitem [{\citenamefont {Kunst}\ \emph {et~al.}(2018)\citenamefont {Kunst}, \citenamefont {Edvardsson}, \citenamefont {Budich},\ and\ \citenamefont {Bergholtz}}]{Kunst-BiorthogonalBulk-BoundarySystems-2018e}%
  \BibitemOpen
  \bibfield  {author} {\bibinfo {author} {\bibfnamefont {F.~K.}\ \bibnamefont {Kunst}}, \bibinfo {author} {\bibfnamefont {E.}~\bibnamefont {Edvardsson}}, \bibinfo {author} {\bibfnamefont {J.~C.}\ \bibnamefont {Budich}},\ and\ \bibinfo {author} {\bibfnamefont {E.~J.}\ \bibnamefont {Bergholtz}},\ }\bibfield  {title} {\bibinfo {title} {Biorthogonal bulk-boundary correspondence in non-hermitian systems},\ }\href {https://doi.org/10.1103/PhysRevLett.121.026808} {\bibfield  {journal} {\bibinfo  {journal} {Phys. Rev. Lett.}\ }\textbf {\bibinfo {volume} {121}},\ \bibinfo {pages} {026808} (\bibinfo {year} {2018})}\BibitemShut {NoStop}%
\bibitem [{\citenamefont {Yokomizo}\ and\ \citenamefont {Murakami}(2019)}]{Yokomizo-Non-BlochBandSystems-2019i}%
  \BibitemOpen
  \bibfield  {author} {\bibinfo {author} {\bibfnamefont {K.}~\bibnamefont {Yokomizo}}\ and\ \bibinfo {author} {\bibfnamefont {S.}~\bibnamefont {Murakami}},\ }\bibfield  {title} {\bibinfo {title} {Non-bloch band theory of non-hermitian systems},\ }\href {https://doi.org/10.1103/PhysRevLett.123.066404} {\bibfield  {journal} {\bibinfo  {journal} {Phys. Rev. Lett.}\ }\textbf {\bibinfo {volume} {123}},\ \bibinfo {pages} {066404} (\bibinfo {year} {2019})}\BibitemShut {NoStop}%
\bibitem [{\citenamefont {Yang}\ \emph {et~al.}(2019)\citenamefont {Yang}, \citenamefont {Zhang}, \citenamefont {Fang},\ and\ \citenamefont {Hu}}]{Yang-Non-HermitianBulk-boundaryTheory-2019u}%
  \BibitemOpen
  \bibfield  {author} {\bibinfo {author} {\bibfnamefont {Z.}~\bibnamefont {Yang}}, \bibinfo {author} {\bibfnamefont {K.}~\bibnamefont {Zhang}}, \bibinfo {author} {\bibfnamefont {C.}~\bibnamefont {Fang}},\ and\ \bibinfo {author} {\bibfnamefont {J.}~\bibnamefont {Hu}},\ }\bibfield  {title} {\bibinfo {title} {Non-hermitian bulk-boundary correspondence and auxiliary generalized brillouin zone theory},\ }\href {https://doi.org/10.1103/PhysRevLett.125.226402} {\bibfield  {journal} {\bibinfo  {journal} {Phys. Rev. Lett.}\ }\bibinfo {series} {125},\ \bibinfo {pages} {226402} (\bibinfo {year} {2019})},\ \Eprint {https://arxiv.org/abs/cond-mat.mes-hall} {arXiv:cond-mat.mes-hall} \BibitemShut {NoStop}%
\bibitem [{\citenamefont {Kawabata}\ \emph {et~al.}(2019)\citenamefont {Kawabata}, \citenamefont {Shiozaki}, \citenamefont {Ueda},\ and\ \citenamefont {Sato}}]{Kawabata-SymmetryAndPhysics-2019h}%
  \BibitemOpen
  \bibfield  {author} {\bibinfo {author} {\bibfnamefont {K.}~\bibnamefont {Kawabata}}, \bibinfo {author} {\bibfnamefont {K.}~\bibnamefont {Shiozaki}}, \bibinfo {author} {\bibfnamefont {M.}~\bibnamefont {Ueda}},\ and\ \bibinfo {author} {\bibfnamefont {M.}~\bibnamefont {Sato}},\ }\bibfield  {title} {\bibinfo {title} {Symmetry and topology in non-hermitian physics},\ }\href {https://doi.org/10.1103/PhysRevX.9.041015} {\bibfield  {journal} {\bibinfo  {journal} {Phys. Rev. X.}\ }\textbf {\bibinfo {volume} {9}},\ \bibinfo {pages} {041015} (\bibinfo {year} {2019})}\BibitemShut {NoStop}%
\bibitem [{\citenamefont {Lee}\ and\ \citenamefont {Thomale}(2019)}]{Lee-AnatomyOfSystems-2019s}%
  \BibitemOpen
  \bibfield  {author} {\bibinfo {author} {\bibfnamefont {C.~H.}\ \bibnamefont {Lee}}\ and\ \bibinfo {author} {\bibfnamefont {R.}~\bibnamefont {Thomale}},\ }\bibfield  {title} {\bibinfo {title} {Anatomy of skin modes and topology in non-hermitian systems},\ }\href {https://doi.org/10.1103/PhysRevB.99.201103} {\bibfield  {journal} {\bibinfo  {journal} {Phys. Rev. B.}\ }\textbf {\bibinfo {volume} {99}},\ \bibinfo {pages} {201103} (\bibinfo {year} {2019})}\BibitemShut {NoStop}%
\bibitem [{\citenamefont {Kunst}\ and\ \citenamefont {Dwivedi}(2019)}]{Kunst-Non-HermitianSystemsPerspective-2019i}%
  \BibitemOpen
  \bibfield  {author} {\bibinfo {author} {\bibfnamefont {F.~K.}\ \bibnamefont {Kunst}}\ and\ \bibinfo {author} {\bibfnamefont {V.}~\bibnamefont {Dwivedi}},\ }\bibfield  {title} {\bibinfo {title} {Non-hermitian systems and topology: A transfer-matrix perspective},\ }\href {https://doi.org/10.1103/PhysRevB.99.245116} {\bibfield  {journal} {\bibinfo  {journal} {Phys. Rev. B}\ }\textbf {\bibinfo {volume} {99}},\ \bibinfo {pages} {245116} (\bibinfo {year} {2019})}\BibitemShut {NoStop}%
\bibitem [{\citenamefont {Song}\ \emph {et~al.}(2019)\citenamefont {Song}, \citenamefont {Yao},\ and\ \citenamefont {Wang}}]{Song-Non-HermitianTopologicalSpace-2019r}%
  \BibitemOpen
  \bibfield  {author} {\bibinfo {author} {\bibfnamefont {F.}~\bibnamefont {Song}}, \bibinfo {author} {\bibfnamefont {S.}~\bibnamefont {Yao}},\ and\ \bibinfo {author} {\bibfnamefont {Z.}~\bibnamefont {Wang}},\ }\bibfield  {title} {\bibinfo {title} {Non-hermitian topological invariants in real space},\ }\href {https://doi.org/10.1103/PhysRevLett.123.246801} {\bibfield  {journal} {\bibinfo  {journal} {Phys. Rev. Lett.}\ }\textbf {\bibinfo {volume} {123}},\ \bibinfo {pages} {246801} (\bibinfo {year} {2019})}\BibitemShut {NoStop}%
\bibitem [{\citenamefont {Zhang}\ \emph {et~al.}(2020)\citenamefont {Zhang}, \citenamefont {Yang},\ and\ \citenamefont {Fang}}]{Zhang-CorrespondenceBetweenSystems-2020p}%
  \BibitemOpen
  \bibfield  {author} {\bibinfo {author} {\bibfnamefont {K.}~\bibnamefont {Zhang}}, \bibinfo {author} {\bibfnamefont {Z.}~\bibnamefont {Yang}},\ and\ \bibinfo {author} {\bibfnamefont {C.}~\bibnamefont {Fang}},\ }\bibfield  {title} {\bibinfo {title} {Correspondence between winding numbers and skin modes in non-hermitian systems},\ }\href {https://doi.org/10.1103/PhysRevLett.125.126402} {\bibfield  {journal} {\bibinfo  {journal} {Phys. Rev. Lett.}\ }\textbf {\bibinfo {volume} {125}},\ \bibinfo {pages} {126402} (\bibinfo {year} {2020})}\BibitemShut {NoStop}%
\bibitem [{\citenamefont {Okuma}\ \emph {et~al.}(2020)\citenamefont {Okuma}, \citenamefont {Kawabata}, \citenamefont {Shiozaki},\ and\ \citenamefont {Sato}}]{Okuma-TopologicalOriginEffects-2020t}%
  \BibitemOpen
  \bibfield  {author} {\bibinfo {author} {\bibfnamefont {N.}~\bibnamefont {Okuma}}, \bibinfo {author} {\bibfnamefont {K.}~\bibnamefont {Kawabata}}, \bibinfo {author} {\bibfnamefont {K.}~\bibnamefont {Shiozaki}},\ and\ \bibinfo {author} {\bibfnamefont {M.}~\bibnamefont {Sato}},\ }\bibfield  {title} {\bibinfo {title} {Topological origin of non-hermitian skin effects},\ }\href {https://doi.org/10.1103/PhysRevLett.124.086801} {\bibfield  {journal} {\bibinfo  {journal} {Phys. Rev. Lett.}\ }\textbf {\bibinfo {volume} {124}},\ \bibinfo {pages} {086801} (\bibinfo {year} {2020})}\BibitemShut {NoStop}%
\bibitem [{\citenamefont {Li}\ \emph {et~al.}(0 12)\citenamefont {Li}, \citenamefont {Lee}, \citenamefont {Mu},\ and\ \citenamefont {Gong}}]{Li-CriticalNon-HermitianEffect-2020k}%
  \BibitemOpen
  \bibfield  {author} {\bibinfo {author} {\bibfnamefont {L.}~\bibnamefont {Li}}, \bibinfo {author} {\bibfnamefont {C.~H.}\ \bibnamefont {Lee}}, \bibinfo {author} {\bibfnamefont {S.}~\bibnamefont {Mu}},\ and\ \bibinfo {author} {\bibfnamefont {J.}~\bibnamefont {Gong}},\ }\bibfield  {title} {\bibinfo {title} {Critical non-hermitian skin effect},\ }\href {https://doi.org/10.1038/s41467-020-18917-4} {\bibfield  {journal} {\bibinfo  {journal} {Nat. Commun.}\ }\textbf {\bibinfo {volume} {11}},\ \bibinfo {pages} {5491} (\bibinfo {year} {2020-12})}\BibitemShut {NoStop}%
\bibitem [{\citenamefont {Kawabata}\ \emph {et~al.}(2020{\natexlab{a}})\citenamefont {Kawabata}, \citenamefont {Sato},\ and\ \citenamefont {Shiozaki}}]{Kawabata-Higher-orderNon-HermitianEffect-2020k}%
  \BibitemOpen
  \bibfield  {author} {\bibinfo {author} {\bibfnamefont {K.}~\bibnamefont {Kawabata}}, \bibinfo {author} {\bibfnamefont {M.}~\bibnamefont {Sato}},\ and\ \bibinfo {author} {\bibfnamefont {K.}~\bibnamefont {Shiozaki}},\ }\bibfield  {title} {\bibinfo {title} {Higher-order non-hermitian skin effect},\ }\href {https://doi.org/10.1103/PhysRevB.102.205118} {\bibfield  {journal} {\bibinfo  {journal} {Phys. Rev. B.}\ }\textbf {\bibinfo {volume} {102}},\ \bibinfo {pages} {205118} (\bibinfo {year} {2020}{\natexlab{a}})}\BibitemShut {NoStop}%
\bibitem [{\citenamefont {Longhi}(2020)}]{Longhi-Non-Bloch-BandCollapseTunneling-2020e}%
  \BibitemOpen
  \bibfield  {author} {\bibinfo {author} {\bibfnamefont {S.}~\bibnamefont {Longhi}},\ }\bibfield  {title} {\bibinfo {title} {Non-bloch-band collapse and chiral zener tunneling},\ }\href {https://doi.org/10.1103/PhysRevLett.124.066602} {\bibfield  {journal} {\bibinfo  {journal} {Phys. Rev. Lett.}\ }\textbf {\bibinfo {volume} {124}},\ \bibinfo {pages} {066602} (\bibinfo {year} {2020})}\BibitemShut {NoStop}%
\bibitem [{\citenamefont {Yi}\ and\ \citenamefont {Yang}(2020)}]{Yi-Non-HermitianSkinEffect-2020e}%
  \BibitemOpen
  \bibfield  {author} {\bibinfo {author} {\bibfnamefont {Y.}~\bibnamefont {Yi}}\ and\ \bibinfo {author} {\bibfnamefont {Z.}~\bibnamefont {Yang}},\ }\bibfield  {title} {\bibinfo {title} {Non-hermitian skin modes induced by on-site dissipations and chiral tunneling effect},\ }\href {https://doi.org/10.1103/PhysRevLett.125.186802} {\bibfield  {journal} {\bibinfo  {journal} {Phys. Rev. Lett.}\ }\textbf {\bibinfo {volume} {125}},\ \bibinfo {pages} {186802} (\bibinfo {year} {2020})}\BibitemShut {NoStop}%
\bibitem [{\citenamefont {Yokomizo}\ and\ \citenamefont {Murakami}(2021{\natexlab{a}})}]{Yokomizo-ScalingRuleEffect-2021u}%
  \BibitemOpen
  \bibfield  {author} {\bibinfo {author} {\bibfnamefont {K.}~\bibnamefont {Yokomizo}}\ and\ \bibinfo {author} {\bibfnamefont {S.}~\bibnamefont {Murakami}},\ }\bibfield  {title} {\bibinfo {title} {Scaling rule for the critical non-hermitian skin effect},\ }\href {https://doi.org/10.1103/PhysRevB.104.165117} {\bibfield  {journal} {\bibinfo  {journal} {Phys. Rev. B.}\ }\textbf {\bibinfo {volume} {104}},\ \bibinfo {pages} {165117} (\bibinfo {year} {2021}{\natexlab{a}})}\BibitemShut {NoStop}%
\bibitem [{\citenamefont {Okugawa}\ \emph {et~al.}(2021)\citenamefont {Okugawa}, \citenamefont {Takahashi},\ and\ \citenamefont {Yokomizo}}]{Okugawa-Non-HermitianBandSymmetry-2021p}%
  \BibitemOpen
  \bibfield  {author} {\bibinfo {author} {\bibfnamefont {R.}~\bibnamefont {Okugawa}}, \bibinfo {author} {\bibfnamefont {R.}~\bibnamefont {Takahashi}},\ and\ \bibinfo {author} {\bibfnamefont {K.}~\bibnamefont {Yokomizo}},\ }\bibfield  {title} {\bibinfo {title} {Non-hermitian band topology with generalized inversion symmetry},\ }\href {https://doi.org/10.1103/PhysRevB.103.205205} {\bibfield  {journal} {\bibinfo  {journal} {Phys. Rev. B.}\ }\textbf {\bibinfo {volume} {103}},\ \bibinfo {pages} {205205} (\bibinfo {year} {2021})}\BibitemShut {NoStop}%
\bibitem [{\citenamefont {Lu}\ \emph {et~al.}(2021)\citenamefont {Lu}, \citenamefont {Zhang},\ and\ \citenamefont {Franz}}]{Lu-MagneticSuppressionEffects-2021y}%
  \BibitemOpen
  \bibfield  {author} {\bibinfo {author} {\bibfnamefont {M.}~\bibnamefont {Lu}}, \bibinfo {author} {\bibfnamefont {X.-X.}\ \bibnamefont {Zhang}},\ and\ \bibinfo {author} {\bibfnamefont {M.}~\bibnamefont {Franz}},\ }\bibfield  {title} {\bibinfo {title} {Magnetic suppression of non-hermitian skin effects},\ }\href {https://doi.org/10.1103/PhysRevLett.127.256402} {\bibfield  {journal} {\bibinfo  {journal} {Phys. Rev. Lett.}\ }\textbf {\bibinfo {volume} {127}},\ \bibinfo {pages} {256402} (\bibinfo {year} {2021})}\BibitemShut {NoStop}%
\bibitem [{\citenamefont {Wu}\ \emph {et~al.}(2022)\citenamefont {Wu}, \citenamefont {Xie}, \citenamefont {Zhou},\ and\ \citenamefont {An}}]{Wu-ConnectionsBetweenSystems-2022u}%
  \BibitemOpen
  \bibfield  {author} {\bibinfo {author} {\bibfnamefont {D.}~\bibnamefont {Wu}}, \bibinfo {author} {\bibfnamefont {J.}~\bibnamefont {Xie}}, \bibinfo {author} {\bibfnamefont {Y.}~\bibnamefont {Zhou}},\ and\ \bibinfo {author} {\bibfnamefont {J.}~\bibnamefont {An}},\ }\bibfield  {title} {\bibinfo {title} {Connections between the open-boundary spectrum and the generalized brillouin zone in non-hermitian systems},\ }\href {https://doi.org/10.1103/PhysRevB.105.045422} {\bibfield  {journal} {\bibinfo  {journal} {Phys. Rev. B.}\ }\textbf {\bibinfo {volume} {105}},\ \bibinfo {pages} {045422} (\bibinfo {year} {2022})}\BibitemShut {NoStop}%
\bibitem [{\citenamefont {Liang}\ \emph {et~al.}(2022)\citenamefont {Liang}, \citenamefont {Xie}, \citenamefont {Dong}, \citenamefont {Li}, \citenamefont {Li}, \citenamefont {Gadway}, \citenamefont {Yi},\ and\ \citenamefont {Yan}}]{Liang-DynamicSignaturesAtoms-2022w}%
  \BibitemOpen
  \bibfield  {author} {\bibinfo {author} {\bibfnamefont {Q.}~\bibnamefont {Liang}}, \bibinfo {author} {\bibfnamefont {D.}~\bibnamefont {Xie}}, \bibinfo {author} {\bibfnamefont {Z.}~\bibnamefont {Dong}}, \bibinfo {author} {\bibfnamefont {H.}~\bibnamefont {Li}}, \bibinfo {author} {\bibfnamefont {H.}~\bibnamefont {Li}}, \bibinfo {author} {\bibfnamefont {B.}~\bibnamefont {Gadway}}, \bibinfo {author} {\bibfnamefont {W.}~\bibnamefont {Yi}},\ and\ \bibinfo {author} {\bibfnamefont {B.}~\bibnamefont {Yan}},\ }\bibfield  {title} {\bibinfo {title} {Dynamic signatures of non-hermitian skin effect and topology in ultracold atoms},\ }\href {https://doi.org/10.1103/PhysRevLett.129.070401} {\bibfield  {journal} {\bibinfo  {journal} {Phys. Rev. Lett.}\ }\textbf {\bibinfo {volume} {129}},\ \bibinfo {pages} {070401} (\bibinfo {year} {2022})}\BibitemShut {NoStop}%
\bibitem [{\citenamefont {Song}\ \emph {et~al.}(2022)\citenamefont {Song}, \citenamefont {Wang},\ and\ \citenamefont {Wang}}]{Song-Non-BlochPTSurprise-2022k}%
  \BibitemOpen
  \bibfield  {author} {\bibinfo {author} {\bibfnamefont {F.}~\bibnamefont {Song}}, \bibinfo {author} {\bibfnamefont {H.-Y.}\ \bibnamefont {Wang}},\ and\ \bibinfo {author} {\bibfnamefont {Z.}~\bibnamefont {Wang}},\ }\bibinfo {title} {Non-bloch {PT} symmetry: Universal threshold and dimensional surprise},\ in\ \href {https://www.worldscientific.com/doi/abs/10.1142/9789811264153_0017} {\emph {\bibinfo {booktitle} {A Festschrift in Honor of the C N Yang Centenary}}}\ (\bibinfo  {publisher} {WORLD SCIENTIFIC},\ \bibinfo {year} {2022})\ pp.\ \bibinfo {pages} {299--311}\BibitemShut {NoStop}%
\bibitem [{\citenamefont {Longhi}(2022)}]{Longhi-Non-HermitianSkinSelf-accelerati-2022s}%
  \BibitemOpen
  \bibfield  {author} {\bibinfo {author} {\bibfnamefont {S.}~\bibnamefont {Longhi}},\ }\bibfield  {title} {\bibinfo {title} {Non-hermitian skin effect and self-acceleration},\ }\href {https://doi.org/10.1103/PhysRevB.105.245143} {\bibfield  {journal} {\bibinfo  {journal} {Phys. Rev. B.}\ }\textbf {\bibinfo {volume} {105}},\ \bibinfo {pages} {245143} (\bibinfo {year} {2022})}\BibitemShut {NoStop}%
\bibitem [{\citenamefont {Gu}\ \emph {et~al.}(2022)\citenamefont {Gu}, \citenamefont {Gao}, \citenamefont {Xue}, \citenamefont {Li}, \citenamefont {Su},\ and\ \citenamefont {Zhu}}]{Gu-TransientNon-HermitianEffect-2022a}%
  \BibitemOpen
  \bibfield  {author} {\bibinfo {author} {\bibfnamefont {Z.}~\bibnamefont {Gu}}, \bibinfo {author} {\bibfnamefont {H.}~\bibnamefont {Gao}}, \bibinfo {author} {\bibfnamefont {H.}~\bibnamefont {Xue}}, \bibinfo {author} {\bibfnamefont {J.}~\bibnamefont {Li}}, \bibinfo {author} {\bibfnamefont {Z.}~\bibnamefont {Su}},\ and\ \bibinfo {author} {\bibfnamefont {J.}~\bibnamefont {Zhu}},\ }\bibfield  {title} {\bibinfo {title} {Transient non-hermitian skin effect},\ }\href {https://doi.org/10.1038/s41467-022-35448-2} {\bibfield  {journal} {\bibinfo  {journal} {Nat. Commun.}\ }\textbf {\bibinfo {volume} {13}},\ \bibinfo {pages} {7668} (\bibinfo {year} {2022})}\BibitemShut {NoStop}%
\bibitem [{\citenamefont {Kawabata}\ \emph {et~al.}(2023)\citenamefont {Kawabata}, \citenamefont {Numasawa},\ and\ \citenamefont {Ryu}}]{Kawabata-EntanglementPhaseEffect-2023q}%
  \BibitemOpen
  \bibfield  {author} {\bibinfo {author} {\bibfnamefont {K.}~\bibnamefont {Kawabata}}, \bibinfo {author} {\bibfnamefont {T.}~\bibnamefont {Numasawa}},\ and\ \bibinfo {author} {\bibfnamefont {S.}~\bibnamefont {Ryu}},\ }\bibfield  {title} {\bibinfo {title} {Entanglement phase transition induced by the non-hermitian skin effect},\ }\href {https://doi.org/10.1103/PhysRevX.13.021007} {\bibfield  {journal} {\bibinfo  {journal} {Phys. Rev. X.}\ }\textbf {\bibinfo {volume} {13}},\ \bibinfo {pages} {021007} (\bibinfo {year} {2023})}\BibitemShut {NoStop}%
\bibitem [{\citenamefont {Li}\ \emph {et~al.}(2023)\citenamefont {Li}, \citenamefont {Trauzettel}, \citenamefont {Neupert},\ and\ \citenamefont {Zhang}}]{Li-EnhancementOfFields-2023n}%
  \BibitemOpen
  \bibfield  {author} {\bibinfo {author} {\bibfnamefont {C.-A.}\ \bibnamefont {Li}}, \bibinfo {author} {\bibfnamefont {B.}~\bibnamefont {Trauzettel}}, \bibinfo {author} {\bibfnamefont {T.}~\bibnamefont {Neupert}},\ and\ \bibinfo {author} {\bibfnamefont {S.-B.}\ \bibnamefont {Zhang}},\ }\bibfield  {title} {\bibinfo {title} {Enhancement of second-order non-hermitian skin effect by magnetic fields},\ }\href {https://doi.org/10.1103/PhysRevLett.131.116601} {\bibfield  {journal} {\bibinfo  {journal} {Phys. Rev. Lett.}\ }\textbf {\bibinfo {volume} {131}},\ \bibinfo {pages} {116601} (\bibinfo {year} {2023})}\BibitemShut {NoStop}%
\bibitem [{\citenamefont {Zhang}\ \emph {et~al.}(2024{\natexlab{a}})\citenamefont {Zhang}, \citenamefont {Yang},\ and\ \citenamefont {Sun}}]{Zhang-EdgeTheoryDimensions-2024x}%
  \BibitemOpen
  \bibfield  {author} {\bibinfo {author} {\bibfnamefont {K.}~\bibnamefont {Zhang}}, \bibinfo {author} {\bibfnamefont {Z.}~\bibnamefont {Yang}},\ and\ \bibinfo {author} {\bibfnamefont {K.}~\bibnamefont {Sun}},\ }\bibfield  {title} {\bibinfo {title} {Edge theory of non-hermitian skin modes in higher dimensions},\ }\href {https://doi.org/10.1103/PhysRevB.109.165127} {\bibfield  {journal} {\bibinfo  {journal} {Phys. Rev. B.}\ }\textbf {\bibinfo {volume} {109}},\ \bibinfo {pages} {165127} (\bibinfo {year} {2024}{\natexlab{a}})}\BibitemShut {NoStop}%
\bibitem [{\citenamefont {Zhang}\ \emph {et~al.}(2024{\natexlab{b}})\citenamefont {Zhang}, \citenamefont {Wang}, \citenamefont {Liu}, \citenamefont {Chen},\ and\ \citenamefont {Jiang}}]{Zhang-HybridSkin-topologicaNumbers-2024r}%
  \BibitemOpen
  \bibfield  {author} {\bibinfo {author} {\bibfnamefont {Y.-L.}\ \bibnamefont {Zhang}}, \bibinfo {author} {\bibfnamefont {L.-W.}\ \bibnamefont {Wang}}, \bibinfo {author} {\bibfnamefont {Y.}~\bibnamefont {Liu}}, \bibinfo {author} {\bibfnamefont {Z.-X.}\ \bibnamefont {Chen}},\ and\ \bibinfo {author} {\bibfnamefont {J.-H.}\ \bibnamefont {Jiang}},\ }\href {http://arxiv.org/abs/2411.07465} {\bibinfo {title} {Hybrid skin-topological effect in non-hermitian checkerboard lattices with large chern numbers}} (\bibinfo {year} {2024}{\natexlab{b}}),\ \Eprint {https://arxiv.org/abs/2411.07465} {arXiv:2411.07465} \BibitemShut {NoStop}%
\bibitem [{\citenamefont {Kawabata}\ \emph {et~al.}(2020{\natexlab{b}})\citenamefont {Kawabata}, \citenamefont {Okuma},\ and\ \citenamefont {Sato}}]{Kawabata-Non-BlochBandClass-2020k}%
  \BibitemOpen
  \bibfield  {author} {\bibinfo {author} {\bibfnamefont {K.}~\bibnamefont {Kawabata}}, \bibinfo {author} {\bibfnamefont {N.}~\bibnamefont {Okuma}},\ and\ \bibinfo {author} {\bibfnamefont {M.}~\bibnamefont {Sato}},\ }\bibfield  {title} {\bibinfo {title} {Non-bloch band theory of non-hermitian hamiltonians in the symplectic class},\ }\href {https://doi.org/10.1103/PhysRevB.101.195147} {\bibfield  {journal} {\bibinfo  {journal} {Phys. Rev. B.}\ }\textbf {\bibinfo {volume} {101}},\ \bibinfo {pages} {195147} (\bibinfo {year} {2020}{\natexlab{b}})}\BibitemShut {NoStop}%
\bibitem [{\citenamefont {Hu}\ and\ \citenamefont {Zhao}(2021)}]{Hu-KnotsAndBands-2021v}%
  \BibitemOpen
  \bibfield  {author} {\bibinfo {author} {\bibfnamefont {H.}~\bibnamefont {Hu}}\ and\ \bibinfo {author} {\bibfnamefont {E.}~\bibnamefont {Zhao}},\ }\bibfield  {title} {\bibinfo {title} {Knots and non-hermitian bloch bands},\ }\href {https://doi.org/10.1103/PhysRevLett.126.010401} {\bibfield  {journal} {\bibinfo  {journal} {Phys. Rev. Lett.}\ }\textbf {\bibinfo {volume} {126}},\ \bibinfo {pages} {010401} (\bibinfo {year} {2021})}\BibitemShut {NoStop}%
\bibitem [{\citenamefont {Yokomizo}\ and\ \citenamefont {Murakami}(2021{\natexlab{b}})}]{Yokomizo-Non-BlochBandSystems-2021g}%
  \BibitemOpen
  \bibfield  {author} {\bibinfo {author} {\bibfnamefont {K.}~\bibnamefont {Yokomizo}}\ and\ \bibinfo {author} {\bibfnamefont {S.}~\bibnamefont {Murakami}},\ }\bibfield  {title} {\bibinfo {title} {Non-bloch band theory in bosonic bogoliubov-de gennes systems},\ }\href {https://doi.org/10.1103/PhysRevB.103.165123} {\bibfield  {journal} {\bibinfo  {journal} {Phys. Rev. B.}\ }\textbf {\bibinfo {volume} {103}},\ \bibinfo {pages} {165123} (\bibinfo {year} {2021}{\natexlab{b}})}\BibitemShut {NoStop}%
\bibitem [{\citenamefont {Li}\ \emph {et~al.}(2021)\citenamefont {Li}, \citenamefont {Sun}, \citenamefont {Zhang},\ and\ \citenamefont {Yi}}]{Li-Non-BlochQuenchDynamics-2021e}%
  \BibitemOpen
  \bibfield  {author} {\bibinfo {author} {\bibfnamefont {T.}~\bibnamefont {Li}}, \bibinfo {author} {\bibfnamefont {J.-Z.}\ \bibnamefont {Sun}}, \bibinfo {author} {\bibfnamefont {Y.-S.}\ \bibnamefont {Zhang}},\ and\ \bibinfo {author} {\bibfnamefont {W.}~\bibnamefont {Yi}},\ }\bibfield  {title} {\bibinfo {title} {Non-bloch quench dynamics},\ }\href {https://doi.org/10.1103/PhysRevResearch.3.023022} {\bibfield  {journal} {\bibinfo  {journal} {Phys. Rev. Res.}\ }\textbf {\bibinfo {volume} {3}},\ \bibinfo {pages} {023022} (\bibinfo {year} {2021})}\BibitemShut {NoStop}%
\bibitem [{\citenamefont {Guo}\ \emph {et~al.}(1 02)\citenamefont {Guo}, \citenamefont {Bao},\ and\ \citenamefont {Tan}}]{Guo-Non-HermitianBulk-boundaryZone-2021c}%
  \BibitemOpen
  \bibfield  {author} {\bibinfo {author} {\bibfnamefont {G.-F.}\ \bibnamefont {Guo}}, \bibinfo {author} {\bibfnamefont {X.-X.}\ \bibnamefont {Bao}},\ and\ \bibinfo {author} {\bibfnamefont {L.}~\bibnamefont {Tan}},\ }\bibfield  {title} {\bibinfo {title} {Non-hermitian bulk-boundary correspondence and singular behaviors of generalized brillouin zone},\ }\href {https://doi.org/10.1088/1367-2630/ac38ce} {\bibfield  {journal} {\bibinfo  {journal} {New J. Phys.}\ }\textbf {\bibinfo {volume} {23}},\ \bibinfo {pages} {123007} (\bibinfo {year} {2021-02})}\BibitemShut {NoStop}%
\bibitem [{\citenamefont {Xue}\ \emph {et~al.}(2021)\citenamefont {Xue}, \citenamefont {Li}, \citenamefont {Hu}, \citenamefont {Song},\ and\ \citenamefont {Wang}}]{Xue-SimpleFormulasTheory-2021y}%
  \BibitemOpen
  \bibfield  {author} {\bibinfo {author} {\bibfnamefont {W.-T.}\ \bibnamefont {Xue}}, \bibinfo {author} {\bibfnamefont {M.-R.}\ \bibnamefont {Li}}, \bibinfo {author} {\bibfnamefont {Y.-M.}\ \bibnamefont {Hu}}, \bibinfo {author} {\bibfnamefont {F.}~\bibnamefont {Song}},\ and\ \bibinfo {author} {\bibfnamefont {Z.}~\bibnamefont {Wang}},\ }\bibfield  {title} {\bibinfo {title} {Simple formulas of directional amplification from non-bloch band theory},\ }\href {https://doi.org/10.1103/physrevb.103.l241408} {\bibfield  {journal} {\bibinfo  {journal} {Phys. Rev. B.}\ }\textbf {\bibinfo {volume} {103}},\ \bibinfo {pages} {L241408} (\bibinfo {year} {2021})}\BibitemShut {NoStop}%
\bibitem [{\citenamefont {Li}\ \emph {et~al.}(2022)\citenamefont {Li}, \citenamefont {Ji}, \citenamefont {Chen}, \citenamefont {Yan},\ and\ \citenamefont {Yang}}]{Li-TopologicalEnergyBands-2022x}%
  \BibitemOpen
  \bibfield  {author} {\bibinfo {author} {\bibfnamefont {Y.}~\bibnamefont {Li}}, \bibinfo {author} {\bibfnamefont {X.}~\bibnamefont {Ji}}, \bibinfo {author} {\bibfnamefont {Y.}~\bibnamefont {Chen}}, \bibinfo {author} {\bibfnamefont {X.}~\bibnamefont {Yan}},\ and\ \bibinfo {author} {\bibfnamefont {X.}~\bibnamefont {Yang}},\ }\bibfield  {title} {\bibinfo {title} {Topological energy braiding of non-bloch bands},\ }\href {https://doi.org/10.1103/PhysRevB.106.195425} {\bibfield  {journal} {\bibinfo  {journal} {Phys. Rev. B.}\ }\textbf {\bibinfo {volume} {106}},\ \bibinfo {pages} {195425} (\bibinfo {year} {2022})}\BibitemShut {NoStop}%
\bibitem [{\citenamefont {Li}\ and\ \citenamefont {Wan}(2022)}]{Li-ExactFormulasSystems-2022p}%
  \BibitemOpen
  \bibfield  {author} {\bibinfo {author} {\bibfnamefont {H.}~\bibnamefont {Li}}\ and\ \bibinfo {author} {\bibfnamefont {S.}~\bibnamefont {Wan}},\ }\bibfield  {title} {\bibinfo {title} {Exact formulas of the end-to-end green's functions in non-hermitian systems},\ }\href {https://doi.org/10.1103/PhysRevB.105.045122} {\bibfield  {journal} {\bibinfo  {journal} {Phys. Rev. B.}\ }\textbf {\bibinfo {volume} {105}},\ \bibinfo {pages} {045122} (\bibinfo {year} {2022})}\BibitemShut {NoStop}%
\bibitem [{\citenamefont {Hu}\ and\ \citenamefont {Wang}(2023)}]{Hu-GreensFunctionsSystems-2023u}%
  \BibitemOpen
  \bibfield  {author} {\bibinfo {author} {\bibfnamefont {Y.-M.}\ \bibnamefont {Hu}}\ and\ \bibinfo {author} {\bibfnamefont {Z.}~\bibnamefont {Wang}},\ }\bibfield  {title} {\bibinfo {title} {Green's functions of multiband non-hermitian systems},\ }\bibfield  {journal} {\bibinfo  {journal} {Phys. Rev. Res.}\ }\href {https://doi.org/10.48550/arXiv.2304.14438} {10.48550/arXiv.2304.14438} (\bibinfo {year} {2023})\BibitemShut {NoStop}%
\bibitem [{\citenamefont {Hu}\ \emph {et~al.}(2023)\citenamefont {Hu}, \citenamefont {Huang}, \citenamefont {Xue},\ and\ \citenamefont {Wang}}]{Hu-Non-BlochBandSystems-2023g}%
  \BibitemOpen
  \bibfield  {author} {\bibinfo {author} {\bibfnamefont {Y.-M.}\ \bibnamefont {Hu}}, \bibinfo {author} {\bibfnamefont {Y.-Q.}\ \bibnamefont {Huang}}, \bibinfo {author} {\bibfnamefont {W.-T.}\ \bibnamefont {Xue}},\ and\ \bibinfo {author} {\bibfnamefont {Z.}~\bibnamefont {Wang}},\ }\bibfield  {title} {\bibinfo {title} {Non-bloch band theory for non-hermitian continuum systems},\ }\bibfield  {journal} {\bibinfo  {journal} {Phys. Rev. B.}\ }\href {https://doi.org/10.48550/arXiv.2310.08572} {10.48550/arXiv.2310.08572} (\bibinfo {year} {2023}),\ \Eprint {https://arxiv.org/abs/cond-mat.mes-hall} {arXiv:cond-mat.mes-hall} \BibitemShut {NoStop}%
\bibitem [{\citenamefont {Liu}\ \emph {et~al.}(2023)\citenamefont {Liu}, \citenamefont {Lu}, \citenamefont {Zhang},\ and\ \citenamefont {Jiang}}]{Liu-ModifiedGeneralizedDisorder-2023n}%
  \BibitemOpen
  \bibfield  {author} {\bibinfo {author} {\bibfnamefont {H.}~\bibnamefont {Liu}}, \bibinfo {author} {\bibfnamefont {M.}~\bibnamefont {Lu}}, \bibinfo {author} {\bibfnamefont {Z.-Q.}\ \bibnamefont {Zhang}},\ and\ \bibinfo {author} {\bibfnamefont {H.}~\bibnamefont {Jiang}},\ }\bibfield  {title} {\bibinfo {title} {Modified generalized brillouin zone theory with on-site disorder},\ }\href {https://doi.org/10.1103/PhysRevB.107.144204} {\bibfield  {journal} {\bibinfo  {journal} {Phys. Rev. B.}\ }\textbf {\bibinfo {volume} {107}},\ \bibinfo {pages} {144204} (\bibinfo {year} {2023})}\BibitemShut {NoStop}%
\bibitem [{\citenamefont {Tai}\ and\ \citenamefont {Lee}(2023)}]{Tai-ZoologyOfTopology-2023l}%
  \BibitemOpen
  \bibfield  {author} {\bibinfo {author} {\bibfnamefont {T.}~\bibnamefont {Tai}}\ and\ \bibinfo {author} {\bibfnamefont {C.~H.}\ \bibnamefont {Lee}},\ }\bibfield  {title} {\bibinfo {title} {Zoology of non-hermitian spectra and their graph topology},\ }\href {https://doi.org/10.1103/PhysRevB.107.L220301} {\bibfield  {journal} {\bibinfo  {journal} {Phys. Rev. B.}\ }\textbf {\bibinfo {volume} {107}},\ \bibinfo {pages} {L220301} (\bibinfo {year} {2023})}\BibitemShut {NoStop}%
\bibitem [{\citenamefont {Matsushima}\ and\ \citenamefont {Yamada}(2024)}]{Matsushima-Non-BlochBandSystems-2024u}%
  \BibitemOpen
  \bibfield  {author} {\bibinfo {author} {\bibfnamefont {K.}~\bibnamefont {Matsushima}}\ and\ \bibinfo {author} {\bibfnamefont {T.}~\bibnamefont {Yamada}},\ }\href {http://arxiv.org/abs/2407.09871} {\bibinfo {title} {Non-bloch band theory for time-modulated discrete mechanical systems}} (\bibinfo {year} {2024}),\ \Eprint {https://arxiv.org/abs/2407.09871} {arXiv:2407.09871} \BibitemShut {NoStop}%
\bibitem [{\citenamefont {Verma}\ and\ \citenamefont {Park}(2024)}]{Verma-Non-BlochBandPhases-2024q}%
  \BibitemOpen
  \bibfield  {author} {\bibinfo {author} {\bibfnamefont {S.}~\bibnamefont {Verma}}\ and\ \bibinfo {author} {\bibfnamefont {M.~J.}\ \bibnamefont {Park}},\ }\href {http://arxiv.org/abs/2405.06240} {\bibinfo {title} {Non-bloch band theory of sub-symmetry-protected topological phases}} (\bibinfo {year} {2024}),\ \Eprint {https://arxiv.org/abs/2405.06240} {arXiv:2405.06240} \BibitemShut {NoStop}%
\bibitem [{\citenamefont {Wang}\ \emph {et~al.}(2024{\natexlab{a}})\citenamefont {Wang}, \citenamefont {Wang},\ and\ \citenamefont {Wang}}]{Wang-Non-BlochSelf-energyFermions-2024f}%
  \BibitemOpen
  \bibfield  {author} {\bibinfo {author} {\bibfnamefont {H.-R.}\ \bibnamefont {Wang}}, \bibinfo {author} {\bibfnamefont {Z.}~\bibnamefont {Wang}},\ and\ \bibinfo {author} {\bibfnamefont {Z.}~\bibnamefont {Wang}},\ }\href {http://arxiv.org/abs/2411.13661} {\bibinfo {title} {Non-bloch self-energy of dissipative interacting fermions}} (\bibinfo {year} {2024}{\natexlab{a}}),\ \Eprint {https://arxiv.org/abs/2411.13661} {arXiv:2411.13661} \BibitemShut {NoStop}%
\bibitem [{\citenamefont {Roy}\ \emph {et~al.}(2024)\citenamefont {Roy}, \citenamefont {Gogoi},\ and\ \citenamefont {Basu}}]{Roy-TopologicalCharacterizatioTheory-2024i}%
  \BibitemOpen
  \bibfield  {author} {\bibinfo {author} {\bibfnamefont {K.}~\bibnamefont {Roy}}, \bibinfo {author} {\bibfnamefont {K.}~\bibnamefont {Gogoi}},\ and\ \bibinfo {author} {\bibfnamefont {S.}~\bibnamefont {Basu}},\ }\href {http://arxiv.org/abs/2410.05427} {\bibinfo {title} {Topological characterization of a non-hermitian ladder via floquet non-bloch theory}} (\bibinfo {year} {2024}),\ \Eprint {https://arxiv.org/abs/2410.05427} {arXiv:2410.05427} \BibitemShut {NoStop}%
\bibitem [{\citenamefont {Yang}\ \emph {et~al.}(2024{\natexlab{a}})\citenamefont {Yang}, \citenamefont {Lu},\ and\ \citenamefont {Lu}}]{Yang-EntangelmentEntropyZone-2024p}%
  \BibitemOpen
  \bibfield  {author} {\bibinfo {author} {\bibfnamefont {Z.}~\bibnamefont {Yang}}, \bibinfo {author} {\bibfnamefont {C.}~\bibnamefont {Lu}},\ and\ \bibinfo {author} {\bibfnamefont {X.}~\bibnamefont {Lu}},\ }\bibfield  {title} {\bibinfo {title} {Entangelment entropy on generalized brillouin zone},\ }\href {http://arxiv.org/abs/2406.15564} {\bibfield  {journal} {\bibinfo  {journal} {Phys. Rev. B.}\ } (\bibinfo {year} {2024}{\natexlab{a}})}\BibitemShut {NoStop}%
\bibitem [{\citenamefont {Fu}\ and\ \citenamefont {Zhang}(2024)}]{Fu-BraidingTopologyBands-2024m}%
  \BibitemOpen
  \bibfield  {author} {\bibinfo {author} {\bibfnamefont {Y.}~\bibnamefont {Fu}}\ and\ \bibinfo {author} {\bibfnamefont {Y.}~\bibnamefont {Zhang}},\ }\bibfield  {title} {\bibinfo {title} {Braiding topology of non-hermitian open-boundary bands},\ }\href {http://arxiv.org/abs/2405.11832} {\bibfield  {journal} {\bibinfo  {journal} {Phys. Rev. B.}\ } (\bibinfo {year} {2024})},\ \Eprint {https://arxiv.org/abs/cond-mat.mes-hall} {arXiv:cond-mat.mes-hall} \BibitemShut {NoStop}%
\bibitem [{\citenamefont {Yang}\ \emph {et~al.}(2024{\natexlab{b}})\citenamefont {Yang}, \citenamefont {Lu},\ and\ \citenamefont {Lu}}]{Yang-EntangelmentEntropyZone-2024g}%
  \BibitemOpen
  \bibfield  {author} {\bibinfo {author} {\bibfnamefont {Z.}~\bibnamefont {Yang}}, \bibinfo {author} {\bibfnamefont {C.}~\bibnamefont {Lu}},\ and\ \bibinfo {author} {\bibfnamefont {X.}~\bibnamefont {Lu}},\ }\bibfield  {title} {\bibinfo {title} {Entangelment entropy on generalized brillouin zone},\ }\bibfield  {journal} {\bibinfo  {journal} {Phys. Rev. B.}\ }\href {https://doi.org/10.48550/arXiv.2406.15564} {10.48550/arXiv.2406.15564} (\bibinfo {year} {2024}{\natexlab{b}})\BibitemShut {NoStop}%
\bibitem [{\citenamefont {Wang}\ and\ \citenamefont {Yan}(2024)}]{Wang-GeneralTheorySystems-2024k}%
  \BibitemOpen
  \bibfield  {author} {\bibinfo {author} {\bibfnamefont {S.-X.}\ \bibnamefont {Wang}}\ and\ \bibinfo {author} {\bibfnamefont {Z.}~\bibnamefont {Yan}},\ }\bibfield  {title} {\bibinfo {title} {General theory for infernal points in non-hermitian systems},\ }\href {https://doi.org/10.1103/physrevb.110.l201104} {\bibfield  {journal} {\bibinfo  {journal} {Phys. Rev. B.}\ }\textbf {\bibinfo {volume} {110}},\ \bibinfo {pages} {L201104} (\bibinfo {year} {2024})}\BibitemShut {NoStop}%
\bibitem [{\citenamefont {Hu}\ \emph {et~al.}(2024)\citenamefont {Hu}, \citenamefont {Wang}, \citenamefont {Wang},\ and\ \citenamefont {Song}}]{Hu-GeometricOriginBreaking-2024t}%
  \BibitemOpen
  \bibfield  {author} {\bibinfo {author} {\bibfnamefont {Y.-M.}\ \bibnamefont {Hu}}, \bibinfo {author} {\bibfnamefont {H.-Y.}\ \bibnamefont {Wang}}, \bibinfo {author} {\bibfnamefont {Z.}~\bibnamefont {Wang}},\ and\ \bibinfo {author} {\bibfnamefont {F.}~\bibnamefont {Song}},\ }\bibfield  {title} {\bibinfo {title} {Geometric origin of non-bloch {P} {T} symmetry breaking},\ }\href {https://doi.org/10.1103/PhysRevLett.132.050402} {\bibfield  {journal} {\bibinfo  {journal} {Phys. Rev. Lett.}\ }\textbf {\bibinfo {volume} {132}},\ \bibinfo {pages} {050402} (\bibinfo {year} {2024})}\BibitemShut {NoStop}%
\bibitem [{\citenamefont {Li}\ \emph {et~al.}()\citenamefont {Li}, \citenamefont {Jiang},\ and\ \citenamefont {Lee}}]{Li-Phase-spaceGeneralizedSystems-2025e}%
  \BibitemOpen
  \bibfield  {author} {\bibinfo {author} {\bibfnamefont {Q.}~\bibnamefont {Li}}, \bibinfo {author} {\bibfnamefont {H.}~\bibnamefont {Jiang}},\ and\ \bibinfo {author} {\bibfnamefont {C.~H.}\ \bibnamefont {Lee}},\ }\href {http://arxiv.org/abs/2501.09785} {\bibinfo {title} {Phase-space generalized brillouin zone for spatially inhomogeneous non-hermitian systems}},\ \Eprint {https://arxiv.org/abs/2501.09785} {arXiv:2501.09785} \BibitemShut {NoStop}%
\bibitem [{\citenamefont {Yao}\ \emph {et~al.}(2018)\citenamefont {Yao}, \citenamefont {Song},\ and\ \citenamefont {Wang}}]{Yao-Non-hermitianChernBands-2018x}%
  \BibitemOpen
  \bibfield  {author} {\bibinfo {author} {\bibfnamefont {S.}~\bibnamefont {Yao}}, \bibinfo {author} {\bibfnamefont {F.}~\bibnamefont {Song}},\ and\ \bibinfo {author} {\bibfnamefont {Z.}~\bibnamefont {Wang}},\ }\bibfield  {title} {\bibinfo {title} {Non-hermitian chern bands},\ }\href {https://doi.org/10.1103/physrevlett.121.136802} {\bibfield  {journal} {\bibinfo  {journal} {Phys. Rev. Lett.}\ }\textbf {\bibinfo {volume} {121}},\ \bibinfo {pages} {136802} (\bibinfo {year} {2018})}\BibitemShut {NoStop}%
\bibitem [{\citenamefont {Liu}\ \emph {et~al.}(2019)\citenamefont {Liu}, \citenamefont {Zhang}, \citenamefont {Ai}, \citenamefont {Gong}, \citenamefont {Kawabata}, \citenamefont {Ueda},\ and\ \citenamefont {Nori}}]{Liu-Second-OrderTopologicalSystems-2019c}%
  \BibitemOpen
  \bibfield  {author} {\bibinfo {author} {\bibfnamefont {T.}~\bibnamefont {Liu}}, \bibinfo {author} {\bibfnamefont {Y.-R.}\ \bibnamefont {Zhang}}, \bibinfo {author} {\bibfnamefont {Q.}~\bibnamefont {Ai}}, \bibinfo {author} {\bibfnamefont {Z.}~\bibnamefont {Gong}}, \bibinfo {author} {\bibfnamefont {K.}~\bibnamefont {Kawabata}}, \bibinfo {author} {\bibfnamefont {M.}~\bibnamefont {Ueda}},\ and\ \bibinfo {author} {\bibfnamefont {F.}~\bibnamefont {Nori}},\ }\bibfield  {title} {\bibinfo {title} {Second-order topological phases in non-hermitian systems},\ }\href {https://doi.org/10.1103/PhysRevLett.122.076801} {\bibfield  {journal} {\bibinfo  {journal} {Phys. Rev. Lett.}\ }\textbf {\bibinfo {volume} {122}},\ \bibinfo {pages} {076801} (\bibinfo {year} {2019})}\BibitemShut {NoStop}%
\bibitem [{\citenamefont {Yokomizo}\ and\ \citenamefont {Murakami}(2023)}]{Yokomizo-Non-BlochBandsSystems-2023i}%
  \BibitemOpen
  \bibfield  {author} {\bibinfo {author} {\bibfnamefont {K.}~\bibnamefont {Yokomizo}}\ and\ \bibinfo {author} {\bibfnamefont {S.}~\bibnamefont {Murakami}},\ }\bibfield  {title} {\bibinfo {title} {Non-bloch bands in two-dimensional non-hermitian systems},\ }\href {https://doi.org/10.1103/PhysRevB.107.195112} {\bibfield  {journal} {\bibinfo  {journal} {Phys. Rev. B.}\ }\textbf {\bibinfo {volume} {107}},\ \bibinfo {pages} {195112} (\bibinfo {year} {2023})}\BibitemShut {NoStop}%
\bibitem [{\citenamefont {Jiang}\ and\ \citenamefont {Lee}(2023)}]{Jiang-DimensionalTransmutationNon-Hermiticity-2023f}%
  \BibitemOpen
  \bibfield  {author} {\bibinfo {author} {\bibfnamefont {H.}~\bibnamefont {Jiang}}\ and\ \bibinfo {author} {\bibfnamefont {C.~H.}\ \bibnamefont {Lee}},\ }\bibfield  {title} {\bibinfo {title} {Dimensional transmutation from non-hermiticity},\ }\href {https://doi.org/10.1103/PhysRevLett.131.076401} {\bibfield  {journal} {\bibinfo  {journal} {Phys. Rev. Lett.}\ }\textbf {\bibinfo {volume} {131}},\ \bibinfo {pages} {076401} (\bibinfo {year} {2023})}\BibitemShut {NoStop}%
\bibitem [{\citenamefont {Xu}\ \emph {et~al.}(2023)\citenamefont {Xu}, \citenamefont {Pang}, \citenamefont {Zhang},\ and\ \citenamefont {Yang}}]{Xu-Two-dimensionalAsymptoticConjecture-2023r}%
  \BibitemOpen
  \bibfield  {author} {\bibinfo {author} {\bibfnamefont {Z.}~\bibnamefont {Xu}}, \bibinfo {author} {\bibfnamefont {B.}~\bibnamefont {Pang}}, \bibinfo {author} {\bibfnamefont {K.}~\bibnamefont {Zhang}},\ and\ \bibinfo {author} {\bibfnamefont {Z.}~\bibnamefont {Yang}},\ }\href {http://arxiv.org/abs/2311.16868} {\bibinfo {title} {Two-dimensional asymptotic generalized brillouin zone conjecture}} (\bibinfo {year} {2023}),\ \Eprint {https://arxiv.org/abs/2311.16868} {arXiv:2311.16868} \BibitemShut {NoStop}%
\bibitem [{\citenamefont {Zhang}\ \emph {et~al.}(2024{\natexlab{c}})\citenamefont {Zhang}, \citenamefont {Shu},\ and\ \citenamefont {Sun}}]{Zhang-AlgebraicNon-HermitianDimensions-2024f}%
  \BibitemOpen
  \bibfield  {author} {\bibinfo {author} {\bibfnamefont {K.}~\bibnamefont {Zhang}}, \bibinfo {author} {\bibfnamefont {C.}~\bibnamefont {Shu}},\ and\ \bibinfo {author} {\bibfnamefont {K.}~\bibnamefont {Sun}},\ }\href {http://arxiv.org/abs/2406.06682} {\bibinfo {title} {Algebraic non-hermitian skin effect and unified non-bloch band theory in arbitrary dimensions}} (\bibinfo {year} {2024}{\natexlab{c}}),\ \Eprint {https://arxiv.org/abs/2406.06682} {arXiv:2406.06682} \BibitemShut {NoStop}%
\bibitem [{\citenamefont {Banerjee}\ \emph {et~al.}()\citenamefont {Banerjee}, \citenamefont {Jaiswal}, \citenamefont {Manjunath},\ and\ \citenamefont {Narayan}}]{Ayan-ATropicalGeometricApproach-2023a}%
  \BibitemOpen
  \bibfield  {author} {\bibinfo {author} {\bibfnamefont {A.}~\bibnamefont {Banerjee}}, \bibinfo {author} {\bibfnamefont {R.}~\bibnamefont {Jaiswal}}, \bibinfo {author} {\bibfnamefont {M.}~\bibnamefont {Manjunath}},\ and\ \bibinfo {author} {\bibfnamefont {A.}~\bibnamefont {Narayan}},\ }\bibfield  {title} {\bibinfo {title} {A tropical geometric approach to exceptional points},\ }\href {https://doi.org/10.1073/pnas.2302572120 (2023)} {\bibfield  {journal} {\bibinfo  {journal} {Proc. Natl. Acad. Sci. U.S.A.}\ }\textbf {\bibinfo {volume} {120}},\ \bibinfo {pages} {e2302572120}}\BibitemShut {NoStop}%
\bibitem [{\citenamefont {Wang}\ \emph {et~al.}(2024{\natexlab{b}})\citenamefont {Wang}, \citenamefont {Song},\ and\ \citenamefont {Wang}}]{Wang-AmoebaFormulationDimensions-2024q}%
  \BibitemOpen
  \bibfield  {author} {\bibinfo {author} {\bibfnamefont {H.-Y.}\ \bibnamefont {Wang}}, \bibinfo {author} {\bibfnamefont {F.}~\bibnamefont {Song}},\ and\ \bibinfo {author} {\bibfnamefont {Z.}~\bibnamefont {Wang}},\ }\bibfield  {title} {\bibinfo {title} {Amoeba formulation of non-bloch band theory in arbitrary dimensions},\ }\href {https://doi.org/10.1103/PhysRevX.14.021011} {\bibfield  {journal} {\bibinfo  {journal} {Phys. Rev. X}\ }\textbf {\bibinfo {volume} {14}},\ \bibinfo {pages} {021011} (\bibinfo {year} {2024}{\natexlab{b}})}\BibitemShut {NoStop}%
\bibitem [{\citenamefont {Wang}(2024)}]{Wang-ConstraintsOfSymmetry-2024w}%
  \BibitemOpen
  \bibfield  {author} {\bibinfo {author} {\bibfnamefont {S.-X.}\ \bibnamefont {Wang}},\ }\bibfield  {title} {\bibinfo {title} {Constraints of internal symmetry on the non-hermitian skin effect and bidirectional skin effect under the action of the hermitian conjugate of time-reversal symmetry},\ }\href {https://doi.org/10.1103/PhysRevB.109.L081108} {\bibfield  {journal} {\bibinfo  {journal} {Phys. Rev. B.}\ }\textbf {\bibinfo {volume} {109}},\ \bibinfo {pages} {L081108} (\bibinfo {year} {2024})}\BibitemShut {NoStop}%
\bibitem [{\citenamefont {Xiong}\ and\ \citenamefont {Hu}(2023)}]{Xiong-GraphMorphologyBands-2023o}%
  \BibitemOpen
  \bibfield  {author} {\bibinfo {author} {\bibfnamefont {Y.}~\bibnamefont {Xiong}}\ and\ \bibinfo {author} {\bibfnamefont {H.}~\bibnamefont {Hu}},\ }\href {http://arxiv.org/abs/2311.14921} {\bibinfo {title} {Graph morphology of non-hermitian bands}} (\bibinfo {year} {2023}),\ \Eprint {https://arxiv.org/abs/2311.14921} {arXiv:2311.14921} \BibitemShut {NoStop}%
\bibitem [{\citenamefont {Xiong}\ \emph {et~al.}(2024)\citenamefont {Xiong}, \citenamefont {Xing},\ and\ \citenamefont {Hu}}]{Xiong-Non-HermitianSkinClassification-2024f}%
  \BibitemOpen
  \bibfield  {author} {\bibinfo {author} {\bibfnamefont {Y.}~\bibnamefont {Xiong}}, \bibinfo {author} {\bibfnamefont {Z.-Y.}\ \bibnamefont {Xing}},\ and\ \bibinfo {author} {\bibfnamefont {H.}~\bibnamefont {Hu}},\ }\href {http://arxiv.org/abs/2407.01296} {\bibinfo {title} {Non-hermitian skin effect in arbitrary dimensions: non-bloch band theory and classification}} (\bibinfo {year} {2024}),\ \Eprint {https://arxiv.org/abs/2407.01296} {arXiv:2407.01296} \BibitemShut {NoStop}%
\bibitem [{\citenamefont {Hu}(2025)}]{Hu-TopologicalOriginSpectra-2025i}%
  \BibitemOpen
  \bibfield  {author} {\bibinfo {author} {\bibfnamefont {H.}~\bibnamefont {Hu}},\ }\bibfield  {title} {\bibinfo {title} {Topological origin of non-hermitian skin effect in higher dimensions and uniform spectra},\ }\href {https://doi.org/10.1016/j.scib.2024.07.022} {\bibfield  {journal} {\bibinfo  {journal} {Sci. Bull. (Beijing)}\ }\textbf {\bibinfo {volume} {70}},\ \bibinfo {pages} {51} (\bibinfo {year} {2025})}\BibitemShut {NoStop}%
\bibitem [{\citenamefont {Fu}\ and\ \citenamefont {Zhang}(2023)}]{Fu-AnatomyOfSystems-2023p}%
  \BibitemOpen
  \bibfield  {author} {\bibinfo {author} {\bibfnamefont {Y.}~\bibnamefont {Fu}}\ and\ \bibinfo {author} {\bibfnamefont {Y.}~\bibnamefont {Zhang}},\ }\bibfield  {title} {\bibinfo {title} {Anatomy of open-boundary bulk in multiband non-hermitian systems},\ }\href {https://doi.org/10.1103/PhysRevB.107.115412} {\bibfield  {journal} {\bibinfo  {journal} {Phys. Rev. B.}\ }\textbf {\bibinfo {volume} {107}},\ \bibinfo {pages} {115412} (\bibinfo {year} {2023})}\BibitemShut {NoStop}%
\bibitem [{\citenamefont {Szegö}(1915)}]{Szego-EinGrenzwertsatzFunktion-1915w}%
  \BibitemOpen
  \bibfield  {author} {\bibinfo {author} {\bibfnamefont {G.}~\bibnamefont {Szegö}},\ }\bibfield  {title} {\bibinfo {title} {Ein grenzwertsatz über die toeplitzschen determinanten einer reellen positiven funktion},\ }\href {https://doi.org/10.1007/bf01458220} {\bibfield  {journal} {\bibinfo  {journal} {Math. Ann.}\ }\textbf {\bibinfo {volume} {76}},\ \bibinfo {pages} {490} (\bibinfo {year} {1915})}\BibitemShut {NoStop}%
\bibitem [{Note1()}]{Note1}%
  \BibitemOpen
  \bibinfo {note} {Since the Ronkin function is piecewise linear and has integer-quantized derivatives, the following equivalent expression exists: $\Phi (E) = Q_E(2)-(Q_E'(2-\epsilon )-Q_E'(0+\epsilon ))$.}\BibitemShut {Stop}%
\bibitem [{\citenamefont {Ronkin}(1974)}]{Ronkin-Introduction}%
  \BibitemOpen
  \bibfield  {author} {\bibinfo {author} {\bibfnamefont {L.~I.}\ \bibnamefont {Ronkin}},\ }\href@noop {} {\emph {\bibinfo {title} {Introduction to the Theory of Entire Functions of Several Variables}}},\ \bibinfo {series} {Translations of Mathematical Monographs}, Vol.~\bibinfo {volume} {44}\ (\bibinfo  {publisher} {American Mathematical Society},\ \bibinfo {year} {1974})\ p.\ \bibinfo {pages} {273}\BibitemShut {NoStop}%
\bibitem [{\citenamefont {Leykam}\ \emph {et~al.}(2017)\citenamefont {Leykam}, \citenamefont {Bliokh}, \citenamefont {Huang}, \citenamefont {Chong},\ and\ \citenamefont {Nori}}]{PhysRevLett.118.040401}%
  \BibitemOpen
  \bibfield  {author} {\bibinfo {author} {\bibfnamefont {D.}~\bibnamefont {Leykam}}, \bibinfo {author} {\bibfnamefont {K.~Y.}\ \bibnamefont {Bliokh}}, \bibinfo {author} {\bibfnamefont {C.}~\bibnamefont {Huang}}, \bibinfo {author} {\bibfnamefont {Y.~D.}\ \bibnamefont {Chong}},\ and\ \bibinfo {author} {\bibfnamefont {F.}~\bibnamefont {Nori}},\ }\bibfield  {title} {\bibinfo {title} {Edge modes, degeneracies, and topological numbers in non-hermitian systems},\ }\href {https://doi.org/10.1103/PhysRevLett.118.040401} {\bibfield  {journal} {\bibinfo  {journal} {Phys. Rev. Lett.}\ }\textbf {\bibinfo {volume} {118}},\ \bibinfo {pages} {040401} (\bibinfo {year} {2017})}\BibitemShut {NoStop}%
\bibitem [{\citenamefont {Shen}\ \emph {et~al.}(2018)\citenamefont {Shen}, \citenamefont {Zhen},\ and\ \citenamefont {Fu}}]{PhysRevLett.120.146402}%
  \BibitemOpen
  \bibfield  {author} {\bibinfo {author} {\bibfnamefont {H.}~\bibnamefont {Shen}}, \bibinfo {author} {\bibfnamefont {B.}~\bibnamefont {Zhen}},\ and\ \bibinfo {author} {\bibfnamefont {L.}~\bibnamefont {Fu}},\ }\bibfield  {title} {\bibinfo {title} {Topological band theory for non-hermitian hamiltonians},\ }\href {https://doi.org/10.1103/PhysRevLett.120.146402} {\bibfield  {journal} {\bibinfo  {journal} {Phys. Rev. Lett.}\ }\textbf {\bibinfo {volume} {120}},\ \bibinfo {pages} {146402} (\bibinfo {year} {2018})}\BibitemShut {NoStop}%
\bibitem [{\citenamefont {Yang}\ \emph {et~al.}(2021)\citenamefont {Yang}, \citenamefont {Schnyder}, \citenamefont {Hu},\ and\ \citenamefont {Chiu}}]{PhysRevLett.126.086401}%
  \BibitemOpen
  \bibfield  {author} {\bibinfo {author} {\bibfnamefont {Z.}~\bibnamefont {Yang}}, \bibinfo {author} {\bibfnamefont {A.~P.}\ \bibnamefont {Schnyder}}, \bibinfo {author} {\bibfnamefont {J.}~\bibnamefont {Hu}},\ and\ \bibinfo {author} {\bibfnamefont {C.-K.}\ \bibnamefont {Chiu}},\ }\bibfield  {title} {\bibinfo {title} {Fermion doubling theorems in two-dimensional non-hermitian systems for fermi points and exceptional points},\ }\href {https://doi.org/10.1103/PhysRevLett.126.086401} {\bibfield  {journal} {\bibinfo  {journal} {Phys. Rev. Lett.}\ }\textbf {\bibinfo {volume} {126}},\ \bibinfo {pages} {086401} (\bibinfo {year} {2021})}\BibitemShut {NoStop}%
\end{thebibliography}%

\end{document}